\documentclass{article}

\usepackage{alltt}
\usepackage{color}
\usepackage{amsmath,amssymb}
\usepackage{mathtools}
\usepackage{multicol}        % used for the two-column index
\usepackage{graphicx}          % Include this line if your
\usepackage{subfigure}          % Include this line if your
\usepackage{framed}          % Include this line if your

\definecolor{green2}{rgb}{0,0.62,0.17}
\definecolor{orange}{rgb}{1,0.5,0.0}
\definecolor{b}{rgb}{0,0,1}
\definecolor{r}{rgb}{1,0,0}

 \newcommand{\R}{\mathbb{R}} 
 \newcommand{\dn}{\mathbf{d}}
 \newcommand{\N}{\mathbb{N}}

 \newcommand{\diag}{\mathrm{diag}}
  
  \newcommand{\rank}{\mathrm{rank}}
  
  \newcommand{\zero}{\mathbf{0}}

\newtheorem{definition}{Definition}
\newtheorem{lemma}{Lemma}
\newtheorem{theorem}{Theorem}
\newtheorem{corollary}{Corollary}
\newtheorem{proposition}{Proposition}
\newtheorem{remark}{Remark}
\newtheorem{example}{Example}
\newtheorem{assumption}{Assumption}

\begin{document}
%\begin{frontmatter}
	\title{Consistent discretization of finite/fixed-time controllers}
	%\thanks[ca]{The corresponding author is Xubin Ping (pingxubin@126.com)}
%	\thanks[fi]{This work was partially supported  by the National Natural Science Foundation of China under Grant 62050410352.}
\author{Andrey Polyakov\footnote{Univ. Lille, Inria, CNRS, UMR 9189  CRIStAL, Centralle Lille, F-59000 Lille, France, andrey.polyakov(denis.efimov)@inria.fr}, Denis Efimov$^*$,
Xubin Ping\footnote{Xidian University, Xi’an, Shaanxi, China,  pingxubin@126.com}}
\date{}

%	\begin{keyword}
%		nonlinear control; sampled-time control; homogeneous systems.
%	\end{keyword}
%\end{frontmatter}
\maketitle

	\begin{abstract}
	The paper proposes an algorithm for a discretization (sampled-time implementation) of a homogeneous control preserving the finite-time and nearly fixed-time stability property of the original (sampling-free) system.  The sampling period is assumed to be constant. Both single-input and multiple-input cases are considered. The robustness (Input-to-State Stability) of the obtained  sampled-time control system is studied as well. Theoretical results are supported by numerical simulations.
\end{abstract}

\section{Introduction}
\label{sec:1}
By definition, the homogeneity  is a dilation symmetry introduced by Leonhard Euler in 18th century as follows:   $f(\lambda x)=\lambda ^{\nu} f(x), \forall \lambda>0$, where the coordinate transformation $x\mapsto \lambda x$ is know today as a standard (or Euler) dilation.
A weighted (generalized) dilation is studied since  1950s. An introduction to stability theory of  weighted homogeneous
Ordinary Differential Equations (ODEs) can be found in  
\cite{Zubov1958:IVM}. Extensions of the homogeneity theory to various finite-dimensional
and infinite-dimensional dynamical models are proposed in \cite{Khomenuk1961:IVM}, \cite{Kawski1991:ACDS}, \cite{Folland1975:AM}, \cite{Polyakov2020:Book}. 
Homogeneous differential equations/inclusions form an important class of control system models  \cite{Rosier1992:SCL}, \cite{Perruquetti_etal2008:TAC}, \cite{Andrieu_etal2008:SIAM_JCO}, \cite{EfimovPerruquetti2010:NOLCOS}, \cite{Polyakov_etal2016:RNC}. They appear as local approximations \cite{Hermes1986:SIAM_JCO} or  set-valued extensions \cite{Levant2005:Aut} of nonlinear systems  and include models of process control \cite{Zimenko_etal2017:Aut},
mechanical systems with frictions \cite{Orlov2005:SIAM_JCO}, fluid dynamics \cite{Polyakov2020:Book}, etc.
Stability and stabilization problems were studied for both standard  \cite{Zubov1964:Book}, \cite{Andreini_etal1988:SCL} and weighted homogeneous  \cite{CoronPraly1991:SCL}, \cite{Hermes1995:SCL}, \cite{Praly1997:CDC}, \cite{SepulchreAeyels1996:MCSS}, 
\cite{Grune2000:SIAM_JCO}, \cite{Nakamura_etal2007:CDC} systems which are the most popular today \cite{Orlov2005:SIAM_JCO}, \cite{Levant2005:Aut}, \cite{Perruquetti_etal2008:TAC},  \cite{Andrieu_etal2008:SIAM_JCO}, \cite{Polyakov_etal2016:RNC}.  A homogeneous model predictive control  is introduced  in \cite{Coron_etal2019:SIAM_JCO}. 

An asymptotically stable homogeneous system is  finite-time stable 
in the case of negative homogeneity degree and nearly fixed-time stable in the case of the positive homogeneity degree (see, e.g. \cite{Nakamura_etal2002:SICE}, \cite{BhatBernstein2005:MCSS}, \cite{Andrieu_etal2008:SIAM_JCO}).  However, the finite/fixed-time stability  
is a fragile property, since an arbitrary small measurement delay or an improper discretization of a finite-time or a fixed-time stable ODE may result in a chattering \cite{AcaryBrogliato2010:SCL},  \cite{Levant2010:TAC} or even in a finite-time blow up \cite{Levant2013:CDC}. Moreover, the explicit discretization (sampled-time implementation) of a finite-time control yields a chattering even if this control law is a continuous function of state \cite{Efimov_etal2017:TAC}, \cite{Josse_etal2021:Aut}.
That is why the discretization issues are very important for practical implementation of finite/fixed-time control/estimation algorithms \cite{AcaryBrogliato2010:SCL}, \cite{Kikuuwe_etal2010:TR}, \cite{LivneLevant2014:Aut},  \cite{Koch_etal2019:TAC},  \cite{Huber_etal2020:TAC}, \cite{BrogliatoPolyakov2021:RNC}, \cite{Michel_etal2021:ECC}, \cite{Hanan_etal2021:CDC}. 

The concept of consistent discertization introduced in \cite{Polyakov_etal2019:SIAM_JCO}
postulates that stability properties of a continuous-time  system  must be  preserved in its discrete-time counterpart (approximation). Consistent  discretizations for stable generalized homogeneous ODEs  were developed in \cite{Polyakov_etal2019:SIAM_JCO}, \cite{Sanchez_etal2021:RNC} based on Lyapunov function theory. Some schemes with state dependent discretization step were given in \cite{Efimov_etal2019:Aut}.  Being efficient for numerical simulations, the mentioned schemes  do not allow a consistent discretization (sampled-time implementation) of \textit{finite-time controllers} in the general case. To the best of authors' knowledge,  such implementations are developed only for first order (\cite{AcaryBrogliato2010:SCL}, \cite{Huber_etal2016:TAC}) and  second order systems (\cite{Huber_etal2020:TAC}, \cite{Brogliato_etal2020:TAC}, \cite{Polyakov_etal2022:CDC}).  This paper presents a consistent discretization for a homogeneous controller designed in \cite{Polyakov_etal2015:Aut}, \cite{Zimenko_etal2020:TAC} for multidimensional linear plants.  It is shown that  the sampled-time implementation of the controller according to the developed scheme  preserves  the finite-time and nearly fixed-time  stability property of the original closed-loop continuous-time system in the disturbance-free case. We also prove an Input-to-State Stability (ISS) of the obtained sampled-time control system with respect to bounded additive perturbations and measurement noises. Algorithms are developed for both single-input and multiple-input models. Numerical simulations show an efficiency of this scheme  for complete rejection of the  so-called \textit{numerical} \textit{chattering} \cite{AcaryBrogliato2010:SCL} caused by a sampled-time implementation of a continuous-time control algorithm.

\textit{Notation}:
$\mathbb{N}$ is the set of natural numbers including $0$; $\R$ is the field of real numbers; $\R_+=\{\alpha\in \R:\alpha>0\}$; $\mathbb{C}$ is the field of complex numbers;
$\zero$ is the zero of a vector space (e.g., the zero vector in $\R^n$);
$I_n\in \R^{n\times n}$ is the identity matrix; $e_i=(0,...,0,1,0,...,0)^{\top}\in \R^{n}$ is the $i$-th element of the canonical Euclidean basis;
$W\succ 0$ denotes positive definiteness of a  symmetric matrix $W=W^{\top}\in \R^{n \times n}$; $\lambda_{\max}(W)$ is a maximal eigenvalue of a symmetric matrix $W$; 
 $\|x\|=\sqrt{x^{\top}Px}$ denotes the weighted Euclidean norm in $\R^n$  with a positive definite matrix $P\succ 0$ specified below in each case when $P$ is not arbitrary; the matrix norm is defined as $\|A\|=\sup_{x\neq 0} \frac{\|Ax\|}{\|x\|}$; $S=\{x\in \R^n: \|x\|=1\}$ is the unit sphere;  $\mathcal{K}$ denotes a class of strictly increasing positive definite continuous functions $[0,+\infty)\mapsto [0,+\infty)$; a function $\sigma \in\mathcal{K}$ is of the class $\mathcal{K}_{\infty}$ if $\sigma(s)\to+\infty$ as $s\to +\infty$; a function 
$\sigma: [0,+\infty)\times[0,+\infty)\mapsto[0,+\infty)$ belongs to the class $\mathcal{KL}$ if the function $s\mapsto \sigma(s,\tau)$ belongs to the class $K$ for any fixed $\tau\in [0,+\infty)$ and the function $\tau\mapsto \sigma(s,\tau)$
is monotonically decreasing to zero for any fixed $s\in[0,+\infty)$;
$L^{\infty}(\R,\R^n)$ is the space of the essentially bounded measurable function $\R\mapsto\R^n$; $\|q\|_{L^{\infty}((a,b),\R^n)}=\mathrm{ess}\sup_{t\in (a,b)} \|q(t)\|$ for $q\in L^{\infty}(\R,\R^n)$; $\ell^{\infty}$ is a space uniformly bounded sequences in $\R^n$; $\diag\{a_1,...,a_n\}\in \R^{n\times n}$ is a diagonal matrix.

\section{Problem Statement}
\label{sec:2}

Let us consider a linear control system\vspace{-2mm}
\begin{equation}	\label{eq:main_system}
	\dot x(t)=Ax(t)+Bu(t), \quad t\in \R_+, \quad x(0)=x_0\in \R^n,\vspace{-2mm}
\end{equation}
where $x(t)\in \R^n$ is the system state, $u(t)\in \R^{m}$ is the control input, $A\!\in\! \R^{n\times n}$, $B\!\in\! \R^{n\times m}$  are known matrices.
\begin{definition} \label{def:cons_discr_control}
	Let  the system \eqref{eq:main_system} with a feedback  $u\in C(\R^n\backslash\{\zero\}, \R^m)$ be globally uniformly 
	finite-time\footnote{A system $\dot x=f(t,x), x(0)=x_0$ is  globally uniformly %\vspace{-3mm}
		\begin{itemize}
			\item \textit{Lyapunov stable} if $\exists \sigma\!\in\! \mathcal{K}_{\infty}$:
			$\|x(t,x_0)\|\!\leq\! \sigma(\|x_0\|), \forall t\!\geq\! 0, \forall x_0\!\in\! \R^n$ and for any solution  $x(t,x_0)$ of the system;%\vspace{-3mm}
			\item 
			\textit{finite-time stable}  if  it is globally uniformly  Lyapunov stable and there exists 
			a locally bounded function $T:\R^n \mapsto [0,+\infty)$ such that  any trajectory of the system vanishes to zero in a finite time:  $\|x(t,x_0)\|\!=\!0,\forall t\!\geq\! T(x_0), \forall x_0\!\in\! \R^n$;%\vspace{-3mm}
			\item \textit{nearly fixed-time stable} if  it is globally uniformly Lyapunov stable and $\forall r\!>\!0, \exists T_r\!>\!0$: $\|x(t,x_0)\|\!<\!r, \forall t\!\geq\! T_r, \forall x_0\!\in\! \R^n$\!. 
	\end{itemize}}  (resp. nearly fixed-time) stable.
	A family of functions $\tilde u_h: \R^n\mapsto \R^m$ parameterized by a scalar $h\!>\!0$ is said to be a \textit{consistent discretization} of  $u$ if 
	\begin{itemize}
		\item \textbf{Consistency of Stability}: the  system \eqref{eq:main_system} with
		\begin{equation}\label{eq:sampled_con}
			\!u(t)\!=\!\tilde u_{h}(x(t_i)), \;\; t\!\in\! [t_i,t_{i+1}), \;\; t_i\!=\!ih, i\!\in \N
		\end{equation}
		is globally  uniformly  finite-time (resp., nearly fixed-time)  stable for any $h>0$; %with  the attraction domain $\{x_0: \|x_0\|\leq \xi(h^{-1})\}$;
		\item \textbf{Control Approximation}:  $ \forall r_1\!>\!0, \forall r_2\!>\!r_1\!,\!\exists \, \omega_{r}\!\in\! \mathcal{K}$ :
		\begin{equation}\label{eq:approx}
			\sup_{r_1\leq \|x\|\leq r_2}\|\tilde u_h(x)-u(x)\| \leq \omega_{r}(h).\vspace{-2mm}
		\end{equation}
	\end{itemize} 
	If the above properties  are fulfilled for all $h\in (0,h_{\max})$ with  $0<h_{\max}+\infty$ then 
	the discretization  is called \textit{conditionally consistent}. 
\end{definition}
The first condition of Definition \ref{def:cons_discr_control} asks  that the sampled-time control system  preserves the stability property of the original system for any fixed sampling period $h>0$. The second condition guarantees that the control $\tilde u_h$ is, indeed, an approximation of $u$, i.e.,  $\tilde u_h(x)\to u(x)$ as $h\to 0^+$ uniformly on compacts  from $\R^{n}\backslash\{\zero\}$.  The origin is excluded since a finite-time stabilizing feedback is always non-smooth or even discontinuous at zero.

The aim of the paper is to develop a consistent discretization  of a generalized homogeneous controllers introduced in  \cite{Polyakov_etal2015:Aut}, \cite{Polyakov_etal2016:RNC}, \cite{Zimenko_etal2020:TAC}.  First, we design a universal control discretization  being a mixture of feedforward/feedback algorithms, which guarantees an exact tracking of the states of the original continuous-time closed-loop system at time instances $t_{nk},k\in \N$.  
Next, we present a consistent (in the sense of the above definition) discretization scheme and study its robustness under the condition:
\begin{assumption}\label{as:1}
	The pair $\{A,B\}$ is controllable, the matrix $A$ is nilpotent and  $m=1$.
\end{assumption}
Recall \cite{Zimenko_etal2020:TAC} that a linear system is generalized homogeneous of \textit{non-zero degree} if and only if $A$ is nilpotent.

Finally, we generalize  both schemes to the multiple-input case assuming that the system can be decomposed into single-input subsystems satisfying Assumption 1.
\begin{assumption}\label{as:2}
	Let us assume that 
	\begin{equation}\label{eq:AandB_multi}
		A\!=\!\!\left[
		\begin{smallmatrix}
			A_1 & * &...& * & * \\
			\zero & A_2 & ...&*& *\\
			... & ... & ... & ...&...\\
			\zero & \zero &.... &A_{m\text{--}1} & * \\
			\zero & \zero &.... &\zero & A_m 
		\end{smallmatrix}
		\right]\!\!,  
		B\!=\!\!\left[
		\begin{smallmatrix}
			B_1 & \zero &...& \zero & \zero \\
			\zero & B_2 & ...&\zero& \zero\\
			... & ... & ... & ...&...\\
			\zero & \zero &... &B_{m\text{--}1} & \zero \\
			\zero & \zero &... &\zero & B_m, 
		\end{smallmatrix}
		\right]\!\!,
	\end{equation}
	where  $A_i\in \R^{n_i\times n_i},B_i\in \R^{n_i}, n_i\geq 1: n_1+n_2+....+n_m=n$ and $*$ denotes (possibly) nonzero blocks. The pairs $\{A_{i}, B_i\}$ are controllable and the matrices $A_i$ are nilpotent, $i=1,...,m$.
\end{assumption}   
If the pair $\{A,B\}$ is controllable and $\rank(B)=m$ then  there exists a coordinate transformation \cite{Luenberger1967:TAC} of $\{A,B\}$ to a canonical form similar to \eqref{eq:AandB_multi} ). Assumption \ref{as:2}  asks that the pair $\{A,B\}$ is controllable, the matrix $A$ is nilpotent, $\rank(B)=m$  and the system admits a transformation to the block form \eqref{eq:AandB_multi}.
\section{Preliminaries: Homogeneous systems}
 \subsection{Linear dilation and homogeneous norm}
 The so-called linear (geometric) dilation \cite[Chapter 6]{Polyakov2020:Book} in $\R^n$ is given by\vspace{-2mm}
 \begin{equation}\label{eq:dilation}
 	\dn(s)=e^{sG_{\dn}}=\sum_{i=0}^{\infty} \tfrac{(sG_{\dn})^i}{i!},\quad s\in \R,\vspace{-2mm}
 \end{equation}
 where $G_{\dn}\in \R^{n\times n}$ is an anti-Hurwitz matrix\footnote{A matrix $G_{\dn}\in\R^{n\times n}$ is  aniti-Hurwitz if $-G_{\dn}$ is Hurwitz.} known as the \textit{generator of linear dilation}. The latter guarantees that $\dn$
 satisfies the limit property, $\|\dn(s)x\|\to 0$ as $s\to -\infty$ and $\|\dn(s)x\|\to+\infty$ as $s\to+\infty$, required for a group $\dn$ to be  a dilation in $\R^n$ (see, {\em e.g.}, \cite{Kawski1991:ACDS}). The linear dilation introduces an alternative norm topology in $\R^n$ by means  the so-called canonical homogeneous norm.
 \begin{definition}\cite{Polyakov2020:Book}\label{def:hom_norm}
 	The function $\|\cdot\|_{\dn} : \R^n \mapsto \R_+$ given by  $\|x\|_{\dn}=0$ for $x=\zero$ and\vspace{-2mm} \begin{equation}\label{eq:hom_norm}
 		\|x\|_{\dn}\!=\!e^{s_x}, \;\;  \text{where}\;  s_x\!\in\! \R: \|\dn(-s_x)x\|\!=\!1,\;\;\; x\!\neq\! \zero\!
 		\vspace{-2mm}
 	\end{equation}
 	is called the \textit{canonical homogeneous norm} in $\R^n$, 
 	where $\dn$ is a linear monotone dilation\footnote{A dilation in $\R^n$ is monotone if for any $x\in \R^n\backslash\{\zero\}$ the function $s\mapsto \|\dn(s)x\|, s\in \R$ is strictly increasing.}\!.
 \end{definition}
 
 Notice that $\|x\|=1$ (resp. $\|x\|\leq 1$) is equivalent to $\|x\|_{\dn}=1$ (resp. $\|x\|_{\dn}\leq 1$). For the uniform dilation $\dn(s)=e^{s}I_n, s\in \R$ we have $\|\cdot\|=\|\cdot\|_\dn$.
 
 \begin{theorem}\cite{Polyakov2018:RNC} \label{thm:hom_norm}
 	If $\dn$ is a  monotone dilation  and $\|x\|=\sqrt{x^{\top}Px}$  with a symmetric  matrix $P\in \R^{n\times n}$ satisfying 
 	$PG_{\dn}+G_{\dn}^{\top}P \succ 0, P\succ 0$
 	then  the canonical homogeneous norm   $\|\cdot\|_{\dn}$ 
 	is continuous on $\R^n$ and smooth on $\R^{n}\backslash\{\zero\}$: 
 	\begin{equation}\label{eq:hom_norm_deriv}
 		\tfrac{\partial \|x\|_{\dn}}{\partial x}\!=\!\tfrac{\|x\|_{\dn}x^{\top}\dn^{\top}\!(-\ln \|x\|_{\dn})P\dn(-\ln \|x\|_{\dn})}{x^{\top}\dn^{\top}\!(-\ln \|x\|_{\dn})PG_{\dn} \dn(-\ln \|x\|_{\dn}) x}, \;\; \forall x\!\neq\!\zero;
 	\end{equation}
 	Moreover, 
 	$
 	\underline\sigma(\|x\|)\leq \|x\|_{\dn}\leq \overline\sigma(\|x\|), \forall x\in \R^n,
 	$ with 
 	\[
 	\underline{\sigma}(r)=\left\{
 	\begin{smallmatrix}
 		r^{1/\alpha} & \text{ if } & r\geq 1,\\
 		r^{1/\beta} & \text{ if } & r<1,
 	\end{smallmatrix}
 	\right. \quad 
 	\overline{\sigma}(r)=\left\{
 	\begin{smallmatrix}
 		r^{1/\beta} & \text{ if } & r\geq 1,\\
 		r^{1/\alpha} & \text{ if } & r<1,
 	\end{smallmatrix}
 	\right.	
 	\]
 	where $\alpha=\!0.5 \lambda_{\max}\!\left(P^{1/2}G_{\dn}P^{-1/2}\!+\!P^{-1/2}G^{\top}_{\dn}P^{1/2}\right)\!>\!0$
 	and \\ {\color{white}1} \quad \quad $\beta=\!0.5\lambda_{\min}\!\left(P^{1/2}G_{\dn}P^{-1/2}\!+\!P^{-1/2}G^{\top}_{\dn}P^{1/2}\right)\!>\!0$.
 \end{theorem}
 
 Below the canonical homogeneous norm is utilized as a Lyapunov function for analysis and control design.   
 \begin{remark}[On computation of $\|\cdot\|_{\dn}$]
 	Since the ca\-nonical homogeneous norm is defined implicitly, a computational algorithm is required for its practical implementation. Issues of numerical estimation of $\|\cdot\|_{\dn}$ are studied in \cite{Polyakov_etal2015:Aut}, 	\cite{Polyakov_etal2016:RNC} based on a bisection method. In \cite[Chapter 8]{Polyakov2020:Book} a scheme for an approximation of $\|\cdot\|_{\dn}$ by an explicit homogeneous function is presented. 
 \end{remark}
 \subsection{Homogeneous continuous-time systems}
 \begin{definition}\cite{Kawski1991:ACDS}
 	A vector field $f:\R^n \mapsto \R^n$ (resp. a function $h:\R^n \mapsto \R$) is said to be $\dn$-homogeneous of 
 	degree $\mu\in \R$  if 
 	$
 	f(\dn(s)x)=e^{\mu s}\dn(s)f(x)$ (resp. $h(\dn(s)x)=e^{\mu s}h(x)$), for all $x\in \R^n$, 
 	$s\in\R$.
 \end{definition}
 If  $f$ is $\dn$-homogeneous of degree $\mu$ then solutions of $\dot x\!=\!f(x)$
 are symmetric \cite{Kawski1991:ACDS}: 
 $x(e^{-\mu s}t,\dn(s)x_0)\!=\!\dn(s)x(t,x_0),$ where $x(t,z)$ denotes a solution with $x(0)\!=\!z$.

 \begin{example}\cite{Zimenko_etal2020:TAC}
 	The linear vector field $x\mapsto Ax$, $A\in \R^{n \times n}$ is $\dn$-homogeneous of the degree $\mu\neq 0$ \; $\Leftrightarrow$ \; $A$ is nilpotent \; $\Leftrightarrow$ \; 
 	$AG_{\dn}=(\mu I_n+G_{\dn})A$.
 \end{example}
 
 The homogeneity degree specifies the convergence rate. 
 \begin{theorem}\cite{BhatBernstein2005:MCSS}, \cite{Nakamura2013:NOLCOS}
 	Let $f:\R^n\mapsto \R^n$ be $\dn$-homogeneous of a degree $\mu\in \R$. If the system $\dot x=f(x)$is asymptotically stable then it is
 	globally  uniformly finite-time (nearly fixed-time) stable for $\mu<0$ ($\mu>0$).
 \end{theorem}
 
 The homogeneous control systems are robust (ISS) with respect to a rather large class of perturbations \cite{Hong2001:Aut}, \cite{Andrieu_etal2008:SIAM_JCO}.
 
 \subsection{Homogeneous stabilization of linear plant}
 
 The following theorem merges  results of   \cite{Polyakov_etal2015:Aut}, \cite{Zimenko_etal2020:TAC}, \cite{Nekhoroshikh_etal2021:CDC}.
 \begin{theorem}\label{thm:hom_control}
 	Let a pair $\{A,B\}$ be controllable. Then\vspace{-2mm}	
 	\begin{itemize}
 		\item[1)] any solution $Y_0\in \R^{m\times n}$, $G_0\in \mathbb{R}^{n\times n}$  of the linear algebraic equation\vspace{-2mm}
 		\begin{equation}\label{eq:Y_0G_0}
 			AG_0-G_0A+BY_0=A, \quad G_0B=\zero\vspace{-2mm}
 		\end{equation}
 		is such that the matrix $G_0-I_n$ is invertible, the matrix $G_{\mathbf{d}}=I_n+\mu G_0$ is anti-Hurwitz for any $\mu\in[-1,1/\tilde n]$, where $\tilde n$ is a minimal natural number such that $\rank[B,AB,...,A^{\tilde n-1}B]=n$,  the matrix $A_0=A+BY_0(G_0-I_n)^{-1}$ satisfies	the identity \vspace{-2mm}
 		\begin{equation}\label{eq:hom_A0}
 			A_0G_{\mathbf{d}}=(G_{\mathbf{d}}+\mu I_n)A_0, \quad G_{\mathbf{d}}B=B;\vspace{-2mm}
 		\end{equation} 		
 		\item[2)]  the linear algebraic system  \vspace{-2mm}
 		\begin{equation}\label{eq:LMI}
 			\!	\!\!\!\!\!\!	\begin{array}{c}
 				A_0X\!+\!XA_0^{\top}\!\!+\!BY\!+\!Y^{\top}B^{\top}\!\!+\!\rho(G_{\mathbf{d}}X\!+\!XG_{\mathbf{d}}^{\top})\!=\!\zero, \\ G_{\mathbf{d}}X+XG_{\mathbf{d}}^{\top}\succ 0, \quad X=X^{\top}\succ 0
 			\end{array}			\!\!\!\!\!\! \vspace{-2mm}
 		\end{equation}
 		has a solution $X\in \mathbb{R}^{n\times n}$, $Y\in \mathbb{R}^{m\times n}$ for any $\rho\in \R_+$; \vspace{1mm}
 		\item[3)] the canonical homogeneous norm $\|\cdot\|_{\mathbf{d}}$ induced by the weighted Euclidean norm $\|x\|=\sqrt{x^{\top} Px}$ with $P=X^{-1}$ is a Lyapunov function of the 
 		system \eqref{eq:main_system} with \vspace{-2mm}
 		\begin{equation}\label{eq:hom_control}
 			u(x)=K_0x+\|x\|_{\dn}^{1+\mu}K\dn(-\ln \|x\|_{\dn})x, \vspace{-2mm}
 		\end{equation}
 		\begin{equation}\label{eq:K}
 			K_{0}=Y_0(G_0-I_n)^{-1}, \quad K=YX^{-1},\vspace{-1mm}
 		\end{equation}
 		where $\dn$ is a dilation generated by $G_{\mathbf{d}}$; moreover,\vspace{-2mm}
 		\begin{equation}\label{eq:LF}
 			\tfrac{d}{dt} \|x\|_{\mathbf{d}} = -\rho \|x\|^{1+\mu}_{\mathbf{d}}, \quad x\neq \zero;\vspace{-2mm}
 		\end{equation}		
 		\item[4)] the feedback law  $u$ given by \eqref{eq:hom_control} is continuously differentiable on $\R^{n}\backslash\{\zero\}$, $u$ is continuous at zero if $\mu>-1$ and 
 		$u$ is discontinuous at zero if $\mu=-1$;
 		
 		\item[5)] the system \eqref{eq:main_system}, \eqref{eq:hom_control} is  $\mathbf{d}$-homogeneous of degree $\mu$.
 	\end{itemize}
 	\vspace{-2mm}
 \end{theorem}
 
 Obviously, the closed-loop system \eqref{eq:main_system},\eqref{eq:hom_control} is uniformly finite-time stable if $\mu<0$ 
 and it is nearly fixed-time stable if $\mu> 0$. For $\mu=0$ the control \eqref{eq:hom_control} becomes  
 $u=K_0x+Kx$.
 Such a control law (under some variations and/or simplifications) has been presented  in the literature as a solution to a  finite-time stabilization problem for linear plants  \cite{Korobov1979:DAN},  \cite{Praly1997:CDC}, \cite{Polyakov_etal2015:Aut}. 
 \begin{remark}\label{rem:Gd_diag}
 	Under Assumption \ref{as:1},  the equation \eqref{eq:Y_0G_0} has a unique solution such that $Y_0=\zero$ (i.e., $A_0=A$) and $\exists J\in \R^{n\times n} : J^{-1}G_{0}J=-\diag\{n-1,...,1,0\}$. 	This follows from the fact then the system \eqref{eq:main_system}, in this case, is equivalent to a controlled integrator chain.
 \end{remark}
 
 A topological equivalence of any stable $\dn$-homogeneous system to a standard homogeneous one \cite{Polyakov2018:RNC} allows an explicit representation of solution for  \eqref{eq:main_system}, \eqref{eq:hom_control} to be derived. \vspace{-2mm}
 \begin{corollary}[Explicit representation of solutions]\label{cor:hom_solution}
 	Under conditions of \; Theorem \ref{thm:hom_control} with $\mu\neq 0$, a solution of the closed-loop system \eqref{eq:main_system}, \eqref{eq:hom_control}
 	is unique and \vspace{-2mm}
 	\begin{equation}\label{eq:hom_sol}
 		x(t+\tau)= Q_{\tau}(\|x(t)\|_{\dn}) x(t),\vspace{-2mm}
 	\end{equation} 
 	where  $\tau,t\geq 0$, $Q_{\tau}(0)=\zero$ and for $r>0$ one has \vspace{-2mm}
 	\begin{equation}
 		\label{eq:Q}
 		Q_{\tau}(r)\!=\!
 		\left\{\!
 		\begin{smallmatrix}
 			e^{G_{\dn}\!\ln r} \hat Q\left(\!\frac{\ln (1+\mu
 				\rho \tau r^{\mu})}{\rho \mu}\!\right)
 			e^{-G_{\dn}\!\ln r} & \text{if} & \frac{1}{r^{\mu}}>-\mu\rho\tau,\\
 			0 & \text{if} & \frac{1}{r^{\mu}}\leq -\mu\rho\tau,
 		\end{smallmatrix}
 		\right. 	\!\!\vspace{-2mm}
 	\end{equation}
 	\begin{equation}\label{eq:hatQ}
 		\hat Q(\hat s)\!=\!e^{
 			-\rho G_{\dn}\hat s}e^{(A+B(K_0+K)+\rho G_{\dn})\hat s}, \quad \hat s\geq 0.
 	\end{equation}
 \end{corollary}
 \vspace{-2mm}The proof of this corollary, as well as the proofs of other main and auxiliary results, are given in the Appendix.
 The matrix-valued function $Q_{nh}(\cdot)$ can be easily computed, since elements of a matrix exponential $e^{sM}$ can always be represented as polynomial functions of $s$, $e^{\rho_i}$, $\cos(\omega_i s)$
 and $\sin(\omega_i s)$, where $\rho_i+\mathbf{i} \omega_i$ are eigenvalues of the matrix $M\in \R^{n\times n}$. %Morever, 
 Moreover, if $A_0=A+BK_0,B, K, G_{\dn}$ satisfy \eqref{eq:LMI} then
 the matrix $X^{1/2}(A_0+BK+\rho G_{\dn})X^{-1/2}$ is skew-symmetric and \vspace{-2mm}
 \begin{equation}\label{eq:R(phi)}
 	e^{(A_0+BK+\rho G_{\dn})\phi}=X^{-1/2} R(\phi) X^{1/2},\vspace{-2mm}
 \end{equation}
 where $R(\phi)$ is a rotation matrix for any $\phi\in \R$, i.e., $R(\phi)R^{\top}(\phi)=R^{\top}(\phi)R(\phi)=I_n.$

 \begin{corollary}[On cascade homogeneous control]\label{cor:cascade}
 	Let Assumption \ref{as:2} be fulfilled. 
 	Let $G_{\dn_i}\in \R^{n_i\times n_i}$, $K_{i}\in \R^{1\times n_i}$, $P_i\in \R^{n_i\times n_i}$ and the control $u_i(x_i)$ with $x_i\in\R^{n_i}$ be defined by Theorem \ref{thm:hom_control}
 	for the pairs $\{A_i,B_i\}$ and some $\mu_i\in \R$, $\rho_i>0$ , respectively.
 	Then the system \eqref{eq:main_system} with the control  $u=(u_1,...,u_m)$ is 
 	globally uniformly finite-time stable if $\mu_i<0$
 	(resp.,   nearly fixed-time  stable if $\mu_i\!>\!0$) for all $i=1,2,...,m$.
 \end{corollary}

\section{Discretization of Homogeneous Control }
\subsection{Single-input case}
Let us represent the system \eqref{eq:main_system} with the sampled-time  control $u(t)=u(t_k)$ for $t=[t_k, t_{k+1})$
in the form:\vspace{-2mm}
\begin{equation}\label{eq:dicrete_model}
	x_{k+1}=A_{h}x_k+B_hu(t_k), \quad k\in \N,\vspace{-2mm}
\end{equation}
where $x_k\!=\!x(t_k), t_k\!=\!kh$, $A_h\!=\!e^{hA}$  and $B_h\!=\!\int^h_0 \!e^{sA}B ds$.
The system \eqref{eq:dicrete_model} can be rewritten as follows:\vspace{-2mm}
\begin{equation}\label{eq:mainsystem_dicr_n}
	x_{k+n}=B_hu(t_{k+n-1})+...+A^{n-1}_hB_hu(t_{k})+A^n_hx_{k}.\vspace{-2mm}
\end{equation}
The controllability  of the pair $\{A,B\}$ implies the controllability of the pair
$\{A_h, B_h\}$ and the invertability of\vspace{-2mm}
\begin{equation}\label{eq:W}
	W_h=[B_h,A_hB_h,....,A^{n-1}_hB_h]\vspace{-2mm}
\end{equation}
(see the formulas \eqref{eq:W_inv}, \eqref{eq:Wh} and Lemma \ref{lem:main} in Appendix).

Let the parameters of a stabilizing homogeneous controller \eqref{eq:hom_control} be designed according to Theorem \ref{thm:hom_control} for some $\mu\neq0$. The case $\mu=0$ is omitted since  the control \eqref{eq:hom_control} is a well-known/studied linear feedback in this case. 
By Corollary \ref{cor:hom_solution}, to track the trajectory of the continuous-time (sampling-free) closed-loop homogeneous system \eqref{eq:main_system}, \eqref{eq:hom_control}, the sampled-time control just has to fulfill the following identity
\begin{equation*}
	Q_{n h}(\|x_k\|_{\dn})x_k\!=\!B_hu(t_{k+n-1})\!+\!...\!+\!A^{n-1}B_hu(t_{k})\!+\!A^n_hx_{k}.
\end{equation*}
Indeed, if a sampled-time control is implemented as 
\begin{equation}\label{eq:hom_consist_full}
	\left[
	\begin{smallmatrix}
		u(t_{k+n-1})\\%u(t_{k+n-2})\\
		...\\u(t_{k})
	\end{smallmatrix}
	\right]=W_h^{-1}\left(Q_{nh}(\|x_k\|_{\dn})- A^n_h\right)x_k,
\end{equation}
then the discrete-time system \eqref{eq:dicrete_model}, \eqref{eq:hom_consist_full} tracks any trajectory of the continuous-time system \eqref{eq:main_system}, \eqref{eq:hom_control} at time instances $t_{kn}$, where $k\in \N$. 
\begin{theorem} \label{thm:con_full} The system  \eqref{eq:main_system} with the sampled-time control \eqref{eq:hom_consist_full} is globally uniformly finite-time stable if $\mu<0$ (nearly fixed-time stable if $\mu>0$).
\end{theorem}
Since $u(t_{k+i-1})$ depends on $x_k=x(t_k)$ but not on $x(t_{k+i-1})$, then the  discretization \eqref{eq:hom_consist_full} of the control \eqref{eq:hom_control} could be useful, for example,  if the control sampling is $n$ times  faster than a measurement sampling. In other cases, the control \eqref{eq:hom_consist_full} is  a certain mixture of  feedforward and feedback algorithms, where the state measurements $x(t_{k+i-1})$ for $i=2,...,n-1$  are simply omitted during  the control implementation.	This could badly impact to a robustness and to a precision of the sampled-time controller.  
To avoid this drawback, let us consider  the static feedback law
\begin{equation}\label{eq:hom_consist_reduced}
	\tilde u_h(x_k)=\tilde K_h(\|x_k\|_{\dn})x_k, \vspace{-2mm}
\end{equation}
\begin{equation}
	\tilde K_h(\|x_k\|_{\dn})=e_n^{\top}W_h^{-1}\left(Q_{nh}(\|x_k\|_{\dn})- A^n_h\right),
\end{equation}
which is obtained from \eqref{eq:hom_consist_full} selecting only $u(t_k)$.  

\begin{proposition}[Approximation property]\label{prop:approx}
	Let $u$ be a homogeneous control \eqref{eq:hom_control} designed by Theorem  \ref{thm:hom_control} under Assumption \ref{as:1}.  Then $\tilde u_h(x) \to u(x)  \text{ as } h\to 0^+$ uniformly 
	on compacts from $\R^n\backslash\{\zero\}$.
\end{proposition}
This proposition, in particular, implies that for a sufficiently small $h>0$ the system  \eqref{eq:dicrete_model}, \eqref{eq:hom_consist_reduced} behaves similarly to the continuous-time system  \eqref{eq:main_system}, \eqref{eq:hom_control} at least on small intervals of time.
Let us denote 
\begin{equation}\label{eq:Lh_Fh}
	L_{ h}=B_{h}e_n^{\top}W_{h}^{-1}, \quad F_{h}=A_{h}-L_{h} A_{h}^n,  \quad h>0, 
\end{equation}
\begin{equation}
	M_h(\|x\|_{\dn})x=(F_h+L_h Q_{nh}(\|x\|_{\dn}))x, \quad x\in \R^n,
\end{equation}
and rewrite the  discrete-time system  \eqref{eq:dicrete_model}, \eqref{eq:hom_consist_reduced} as follows\vspace{-2mm}
\begin{equation}\label{eq:disc_time_z}
	x_{k+1}=z_h(x_k):=M_h(\|x_k\|_{\dn})x_k.\vspace{-2mm}
\end{equation}

\begin{lemma}[Homogeneity of discretization]\label{lem:hom_z}
	The system  \eqref{eq:dicrete_model}, \eqref{eq:hom_consist_reduced}
	is $\dn$-homogeneous as follows :\vspace{-2mm}
	\begin{equation}
		z_h(\dn(s)x)=	\dn(s)z_{e^{\mu s} h}\!\left(x\right),\label{eq:hom_z}\vspace{-2mm}
	\end{equation}
	\begin{equation}
		\tilde u_h(\dn(s)x)=e^{s(1+\mu)}\tilde u_{e^{\mu s}h}(x), \label{eq:hom_tilde_u}
	\end{equation}
	for all $s\in \R$, for all $h>0$ and for all $x\in \R^n$
\end{lemma}

The dilation symmetry established by Lemma \ref{lem:hom_z} guarantees that a global asymptotic stability of the discrete-time system \eqref{eq:dicrete_model}, \eqref{eq:hom_consist_reduced} for some $h=\hat h>0$ is equivalent to the global asymptotic stability of this system for any $h>0$. For simplicity,  we select \vspace{-2mm}
\begin{equation}\label{eq:hat_h}
	\hat h:=(|\mu|\rho n)^{-1}.\vspace{-2mm}
\end{equation}
As shown below, the key feature of the proposed control discretization is the nilpotence of the matrix $F_h$.
Together with the properties of $Q_{nh}(\|x_k\|_{\dn})x_k$, this allows the controller \eqref{eq:hom_consist_reduced} to preserve stability properties of the original system. 
\begin{lemma}\label{lem:hom_con_discr}
	Let $u$ be a homogeneous control \eqref{eq:hom_control} designed by Theorem  \ref{thm:hom_control} under Assumption \ref{as:1}.  Then 
	the closed-loop discrete-time system  \eqref{eq:dicrete_model}, \eqref{eq:hom_consist_reduced}
	is\vspace{-2mm}
	\begin{itemize}
		\item[1)] locally uniformly finite-time stable for $\mu\!<\!0$ and \vspace{-2mm}
		\[
		\forall x_0\!\in\! \R^n: \left\|x_0\right\|_{\dn}\! \leq\!  \underline{r}^-\!(\hat h/h)^{1/\mu} \;\; \Rightarrow\;\;  x_k\!=\!\zero,\; \forall k\!\geq\! n, \vspace{-2mm}
		\]
		where $\|x\|_{\dn}$ is the canonical homogeneous norm induced by the  weighted Euclidean norm $\|x\|=\sqrt{x^{\top} Px}$ with $P=X^{-1}$ and 
		$$	
		\underline r^{\text{--}}\!>\!0:  \!\! \max\limits_{i\in\{1,...,n\}}\!\|F^{i-1}_{\hat h} \dn(\ln \underline r^{\text{--}} )\|\! < \!1;\!\!
		$$
		\item[2)] globally practically\footnote{\textit{Practical} finite-time and fixed-time stability is introduced using the same definitions by replacing a norm (distance to $\zero$) with a distance to a set being a neighborhood of zero.} 
		finite-time stable for $\mu<0$  and  the set 
		$$
		\Omega^-=\left\{x\in \R^n: \|x\|_{\dn}\leq  \overline{r}^- (\hat h/h)^{1/\mu}\right\}$$ is invariant and	
		finite-time stable for some $ \overline r^-\geq \underline r^-$;
		
		\item[3)] globally practically fixed-time stable for  $\mu>0$: \vspace{-2mm}
		\begin{equation*}
			\|x_k\|_{\dn}\leq \overline r^+ (\hat h/h)^{1/\mu}, \quad \forall k\geq n\vspace{-2mm}
		\end{equation*}
		for all $x_0\in \R^n$, where 
		$ \overline r^+\!>\!\max\limits_{\|v_i\|\leq 1} \left\|\sum_{i=1}^{n} F_{\hat h}^{i-1}L_{\hat h}v_i\right\|_{\dn};$
		\item[4)]  locally asymptotically stable for $\mu>0$ and  the set $$\Omega^+=\left\{x\!\in \!\R^n: \|x\|_{\dn}\!\leq\! \underline r^+ (\hat h/h)^{1/\mu} \right\}
		$$ is an invariant attraction domain for some $\underline r^+\in (0, \overline r^+]$.
	\end{itemize}
\end{lemma}

The latter lemma proves that the discretization \eqref{eq:hom_consist_reduced} of the controller \eqref{eq:hom_control}, indeed preserves a stability property of the original system at least locally. The discrete-time system with $h=\hat h$ behaves similarly to the continuous-time system for $\|x_k\|_{\dn}<\underline r^{\pm}$ and $\|x_k\|_{\dn}>\overline r^{\pm}$. If 
the set $\tilde \Omega^{\pm}=\{x: \underline r^{\pm}<\|x\|_{\dn}<\overline r^{\pm} \}$ does not contain an invariant set of the discrete-time system 
then the discretization is  globally consistent.  

Let us consider  a family of  mappings $\Theta_k: (0,+\infty)\times S\mapsto \R^{n\times n}$ defined recursively as follows: $\Theta_0(\delta ,v)\!=\! I_n$ and  \vspace{-2mm}
\begin{equation}\label{eq:Theta}
	\quad 
	\Theta_{k+1}(\delta, v)\!=\!M_{\delta \hat h}(\|\Theta_k(\delta,v)v\|_{\dn})\Theta_k(\delta,v), \;\; k\!\in\! \N,\!\!\vspace{-2mm}
\end{equation}
where  $\delta\in \R, v\in S$, the parameter $\hat h>0$ is defined by \eqref{eq:hat_h} and $S=\{x\in \R^n: \|x\|=1\}$ is the unit sphere.
\begin{lemma}\label{lem:sol_discr}
	Any solution of the discrete-time system \eqref{eq:dicrete_model}, \eqref{eq:hom_consist_reduced} with $h\!=\!\hat h$ and $x_0\!\neq\! \zero$ admits the representation\vspace{-2mm}
	\begin{equation}\label{eq:sol_hom_disc}
		x_k=\dn(\ln \|x_0\|_{\dn}) \Theta_k\left(\|x_0\|_{\dn}^{\mu},v_0\right)v_0, \vspace{-2mm}
	\end{equation}
	where $v_0=\dn(-\ln \|x_0\|_{\dn})x_0\in S$.
\end{lemma}
The following theorem presents a necessary and sufficient condition of  the consistency of the discretization  \eqref{eq:hom_consist_reduced} for the controller \eqref{eq:hom_control}. 
\begin{theorem} [Consistent discretization] \label{thm:consistency}Let $u$ be a homogeneous control \eqref{eq:hom_control} designed by Theorem  \ref{thm:hom_control}  under Assumption \ref{as:1} for $\mu\neq 0$.  Then $\tilde u_h$ given by \eqref{eq:hom_consist_reduced} is a consistent discretization of the control $u$ \textbf{if an only if} there exists $k^*\geq 1$ such that 
	\begin{equation}\label{eq:Theta<1}
		\|\Theta_{k^*}(\delta,v) v\|_{\dn}<1, \forall \delta\in (0,r^*], \quad \forall v\in S,
	\end{equation}
	where $r^*=(\underline{r}^-)^{\mu}$ if $\mu<0$ and $r^*=(\overline{r}^+)^{\mu}$ if $\mu>0$.
\end{theorem}
Therefore,  the discrete-time system \eqref{eq:dicrete_model}, \eqref{eq:hom_consist_reduced} is uniformly finite-time(or nearly fixed-time) stable if and only if  the canonical homogeneous norm $\|\cdot\|_{\dn}$ is {a kind of} a homogeneous Lyapunov function. Indeed, \eqref{eq:sol_hom_disc} and \eqref{eq:Theta<1} simply means that
$\|x_{k^*}\|_{\dn}\!<\!\|x_0\|_{\dn}$, $\forall x_0\!:\!\|x_0\|_{\dn}^{\mu}\!\in\! (0,r^*)$.\!\!\!\!

\begin{remark}[On feasibility of condition \eqref{eq:Theta<1}]\label{rem:feas}
	For $k^*=1$ the condition \eqref{eq:Theta<1} is equivalent to the nonlinearly parameterized matrix inequality\vspace{-2mm}
	\begin{equation}\label{eq:stab_cond}
		M_{\delta \hat h}(1)^{\!\top}X^{-1}M_{\delta \hat h}(1)^{\!\top}<\!X^{-1}, \;\;\forall \delta\!\in\! (0, r^*],\vspace{-2mm}
	\end{equation}	
	which can  be checked numerically on a sufficiently dense grid in $(0, r^*]$ because of  a continuous dependence of $M_{\delta \hat h}(1)$ on the parameter $\delta$. Denoting 
	$\Delta(\delta)=X^{-1}-M_{\delta \hat h}^{\top}(1)X^{-1}M_{\delta \hat h}(1))$
	we conclude that the condition \eqref{eq:stab_cond} is fulfilled if $\lambda_{\min}(\Delta (\delta))>0$ for all $\delta\in (0, r^*]$. For example,  for  $n=2$, $\mu=-1,\rho=2$,$A=\left[\begin{smallmatrix} 0 & 1\\ 0 & 0\end{smallmatrix}\right], B=\left[\begin{smallmatrix} 0 \\ 1 \end{smallmatrix}\right]$  Theorem \ref{thm:hom_control} gives $G_{\dn}=\diag\{2,1\}$, $K=YX^{-1}$ with
	$
	X=x_{11}\tilde X,$  $x_{11}>0,$ $\tilde X=\left[
	\begin{smallmatrix}
		1  & -\rho(1-\mu)\\
		- \rho(1-\mu)	& 7(2\text{--}\mu)^2\!\rho^2\!(1\text{--}\mu)/8
	\end{smallmatrix}\right],
	$
	$
	Y=\rho^2(2-\mu)(1-\mu)x_{11}\left[ \begin{smallmatrix}\frac{8-7(2\text{--}\mu)}{8}& \frac{-  7\rho(2\text{--}\mu))}{8} \end{smallmatrix}\right].
	$
	In this case, we have $r^*=1$ and Figure \ref{fig:W} depicts the evolution of the function $\delta \mapsto \lambda_{\min}(\Delta (\delta))$, which confirms that 
	\eqref{eq:stab_cond} is fulfilled.
	\begin{figure}\centering
		\includegraphics[height=40mm,width=50mm, bb=0 0 30cm 30cm]{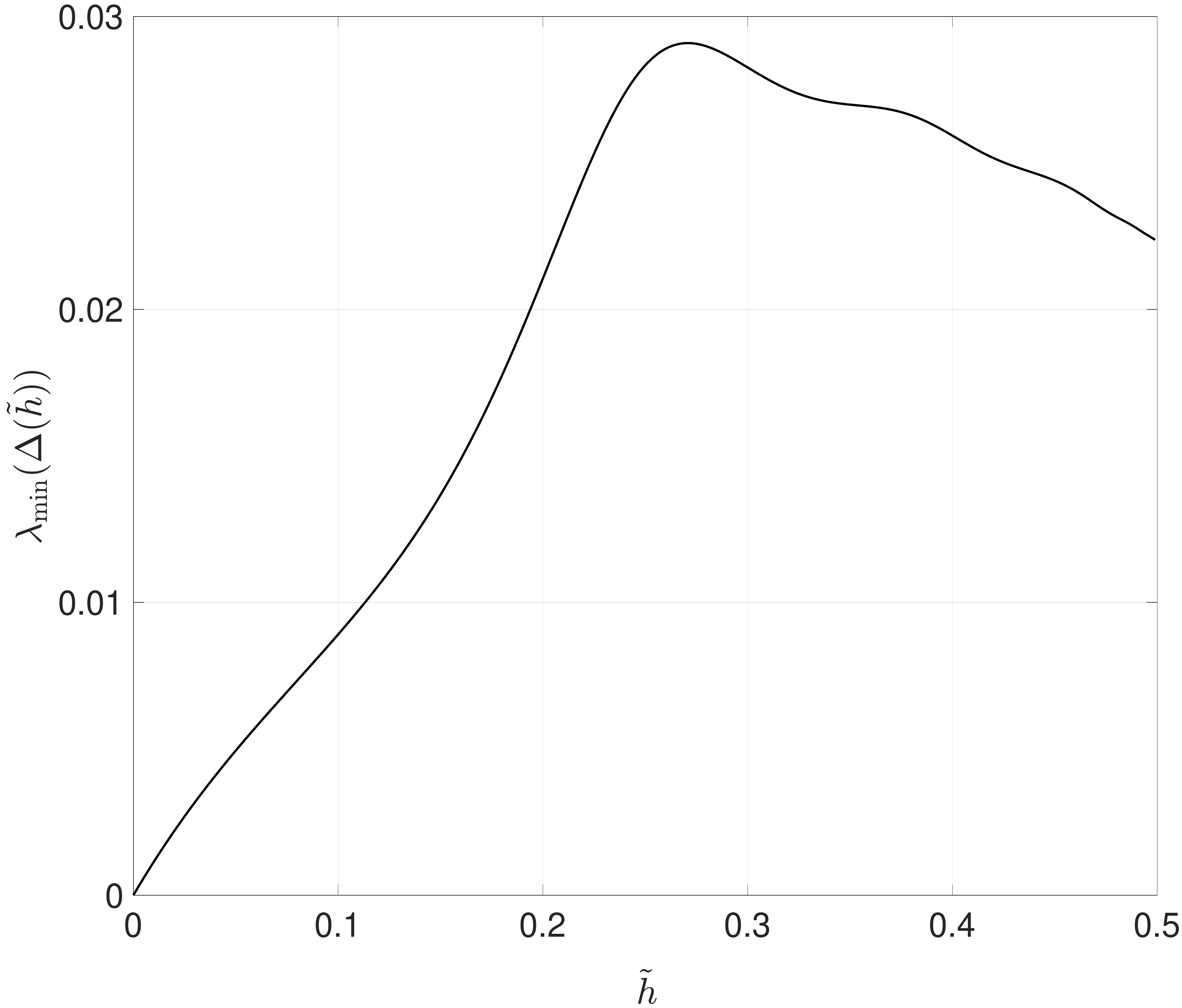}
		\caption{The minimal eigenvalue of the matrix $\Delta(\delta)$ for $\mu\!=\!-1/2, \rho\!=\!2$}
		\label{fig:W}
	\end{figure} For $k^*>1$ a similar (but a bit more complicated)  numerical procedure can be developed.
\end{remark}

\subsection{Robustness analysis}

It is well known \cite{Hong2001:Aut}, \cite{Andrieu_etal2008:SIAM_JCO} that homogeneous systems are Input-to-State Stable (ISS) with respect to sufficiently large class of perturbations. Recall \cite{Sontag1989:TAC} that a system \vspace{-2mm}
\begin{equation}
	\dot x=f(t,x,q), \quad t>t_0 \vspace{-2mm}
\end{equation}
is practically ISS with respect to $q\in L^{\infty}(\R,\R^l)$ if there exist, $c>0$, $\xi\in \mathcal{KL}$ and $\gamma \in \mathcal{K}$ such that \vspace{-2mm}
\begin{equation}\label{eq:pISS}
	\|x(t)\|\leq c+\xi(\|x_0\|,t-t_0)+\gamma(\|q\|_{L^{\infty}((t_0,t),\R^l)}).\vspace{-2mm}
\end{equation}
If $c=0$ then the system is ISS. Local ISS restricts additionally the set of initial conditions $\|x_0\|\leq r_{x}$ and/or the maximal magnitude of the input $\|q\|\leq r_{q}$.

Let us consider the system  \vspace{-2mm}
\begin{equation}\label{eq:system_q0}
	\begin{split}\dot x(t)&=Ax(t)+Bu(t)+q(t), \quad  t>0, \\
		u(t)&= u(t_k),\quad\quad \quad \quad \quad \quad  \quad\;\;\; t\in [t_k,t_{k+1}),
	\end{split}
\end{equation}
where $u(t_k)$ is given by \eqref{eq:hom_consist_full}
and   $q\in L^{\infty}(\R,\R^{n})$ is an exogenous input.
\begin{theorem}\label{thm:ISS0} 
	The system \eqref{eq:system_q0} is 1) locally ISS; 2) practically fixed-time stable if $\mu>0$; 3) ISS  if $\mu>-\beta$, where $\beta>0$ is defined in Theorem \ref{thm:hom_norm}.
\end{theorem}
The ISS can be established for consistent discretization of homogeneous controller.	Let us consider the system  \vspace{-2mm}
\begin{equation}\label{eq:system_q}
	\begin{split}\dot x(t)&=Ax(t)+Bu(t)+q_{p}(t), \quad  t>0, \\
		u(t)&=\tilde u_h(x(t_k)+q_{m}(t_k)), \quad\quad\; t\in [t_k,t_{k+1})\vspace{-2mm}
	\end{split}
\end{equation}
where $\tilde u_h$  given by  \eqref{eq:hom_consist_reduced} is a discretization of  
a control \eqref{eq:hom_control} designed by Theorem  \ref{thm:hom_control} under Assumption \ref{as:1} and 
the exogenous input  $q=(q_p^{\top},q_m^{\top})^{\top} \in L^{\infty}(\R,\R^{2n})$ is such that  $\{q_m(t_k)\}\in \ell^{\infty}$.  Here $q_p$ models an external  perturbation and $q_m$ is a measurement noise. 
\begin{theorem}\label{thm:ISS} 
	The system \eqref{eq:system_q} is 1)
	locally ISS; 2)  practically fixed-time stable if $\mu>0$; 3)  practically  ISS if  $\mu>-\beta$;  4) ISS  if $\mu>-\beta$ and the unperturbed system $(q\!=\!\zero)$ is globally asymptotically stable.
\end{theorem}

Notice that  $q$ may contain an output of another system. In this case,  a  stability analysis of a cascade system can be based on ISS \cite{Sontag1989:TAC}, \cite{ChailletAngeli2014:TAC}.

\subsection{Multiple-input case}

Let the control $u$ be designed using Corollary \ref{cor:cascade} under Assumption \ref{as:2}. Since $A_i$ is nilpotent then, as before, $K_0=\zero$ in Theorem \ref{thm:hom_control} and in Corollary \ref{cor:hom_solution}.
Let $Q_{n_ih}(\cdot)$ be defined by the formula \eqref{eq:Q} for $A_i,B_i,K_i, G_{\dn_i}, P_i, \mu_i$, $i=1,...,m$.
Let us denote\vspace{-2mm}
\begin{equation}\label{eq:W_MIMO}
	W_{h,i}=[B_{h,i},A_{h,i}B_{h,i},....,A^{n-1}_{h,i}B_{h,i}],\vspace{-2mm}
\end{equation}
$
A_{h,i}=e^{hA_i},  B_{h,i}=\int^h_0 e^{\tau A_i} d\tau B,
$
and introduce the following discretization of the controllers $u_i$:\vspace{-2mm}
\begin{equation}\label{eq:hom_consist_full_MIMO}
	\left[
	\begin{smallmatrix}
		u_i(t_{k\text{+}n\text{--}1})\\%u_i(t_{k+n-2})\\
		...\\u_i(t_{k})
	\end{smallmatrix}
	\right]\!\!=\!W_{h,i}^{-1}\!\left(Q_{n_ih}(\|x_{i}(t_k)\|_{\dn})\!-\! A^n_{h,i}\right)\!x_{i}(t_k).\!\!\vspace{-2mm}
\end{equation}
\begin{corollary} \label{cor:con_full_MIMO}Under Assumption \ref{as:2}, the system \eqref{eq:dicrete_model} with the control \eqref{eq:hom_consist_full_MIMO} is globally uniformly 
	finite-time stable if $\mu_i<0$ (nearly fixed-time stable if $\mu_i>0$),  $\forall i=1,...,m$.
\end{corollary}
Similarly to the single input case, a consistent discretization   of the multiple-input control system is  designed as
\begin{equation}\label{eq:hom_consist_reduced_MIMO}
	u_i(x_{k,i})=\tilde K_i(\|x_{k,i}\|_{\dn_i})x_{k,i},
\end{equation}
\[
\tilde K_i(\|x_{k,i}\|_{\dn_i})=e_{n_i}^{\top}W_{h,i}^{-1}\left(Q_{nh,i}(\|x_{k,i}\|_{\dn_i})\!-\! A^n_{h,i}\right).
\]
\begin{corollary}\label{cor:con_reduced_MIMO}
	Let a control  law $u$ be designed by Corollary \ref{cor:cascade} under Assumption \ref{as:2}. 
	Then  \eqref{eq:hom_consist_reduced_MIMO} is a consistent discretization of $u$ provided that conditions of Theorem \ref{thm:consistency} are fulfilled  for  $\mu_i\!<\!0$, $\forall i\!=\!1,...,m$ or $\mu_i\!>\!0$, $\forall i=1,...,m$.
\end{corollary}

\section{Numerical Examples}
\subsection{Single-input system} 
\begin{figure}[h!]
	\centering
	\includegraphics[width=41mm,height=27mm,  bb=0 0 35cm 30cm]{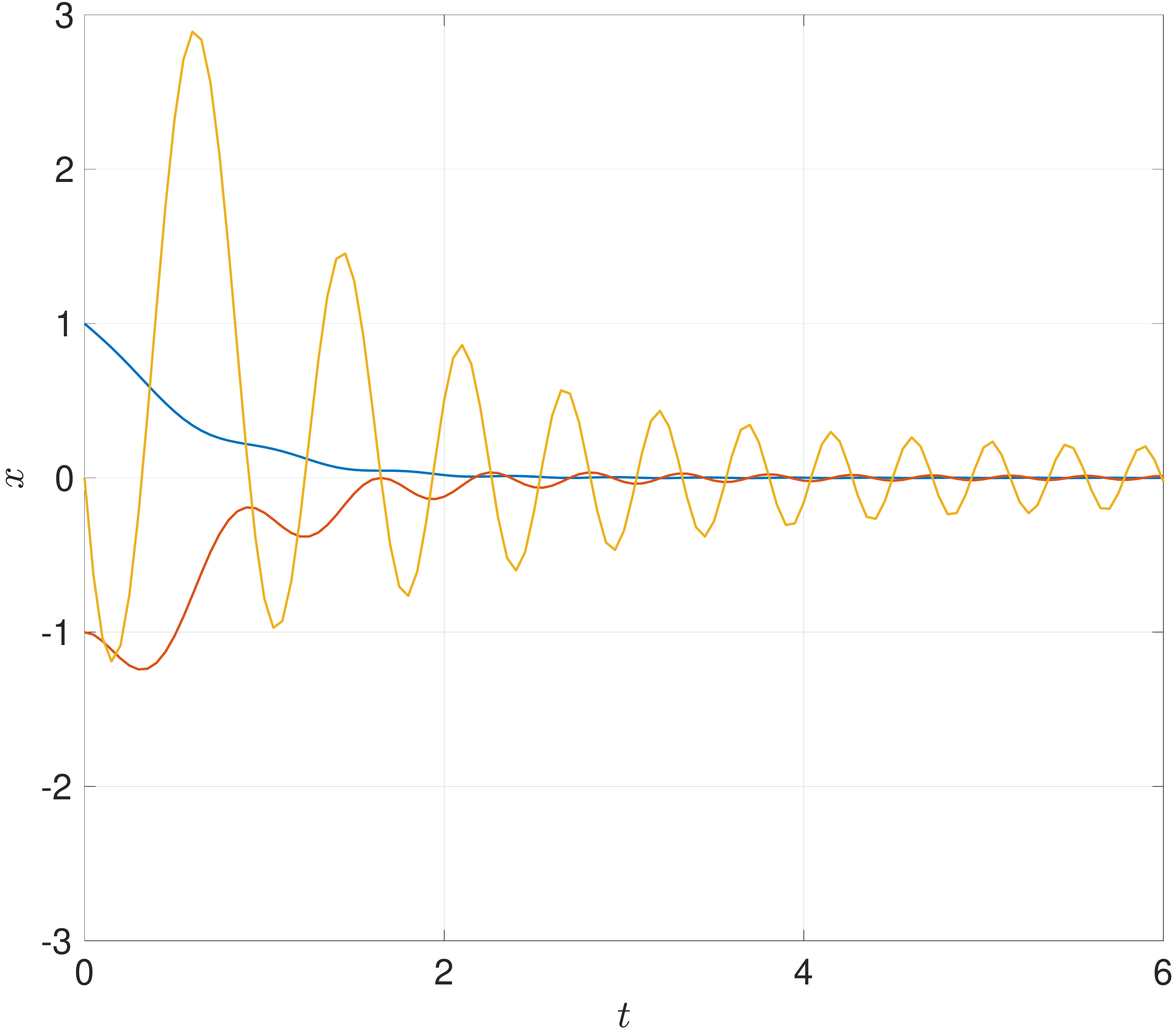}
	\includegraphics[width=41mm,height=27mm, bb=0 0 35cm 30cm]{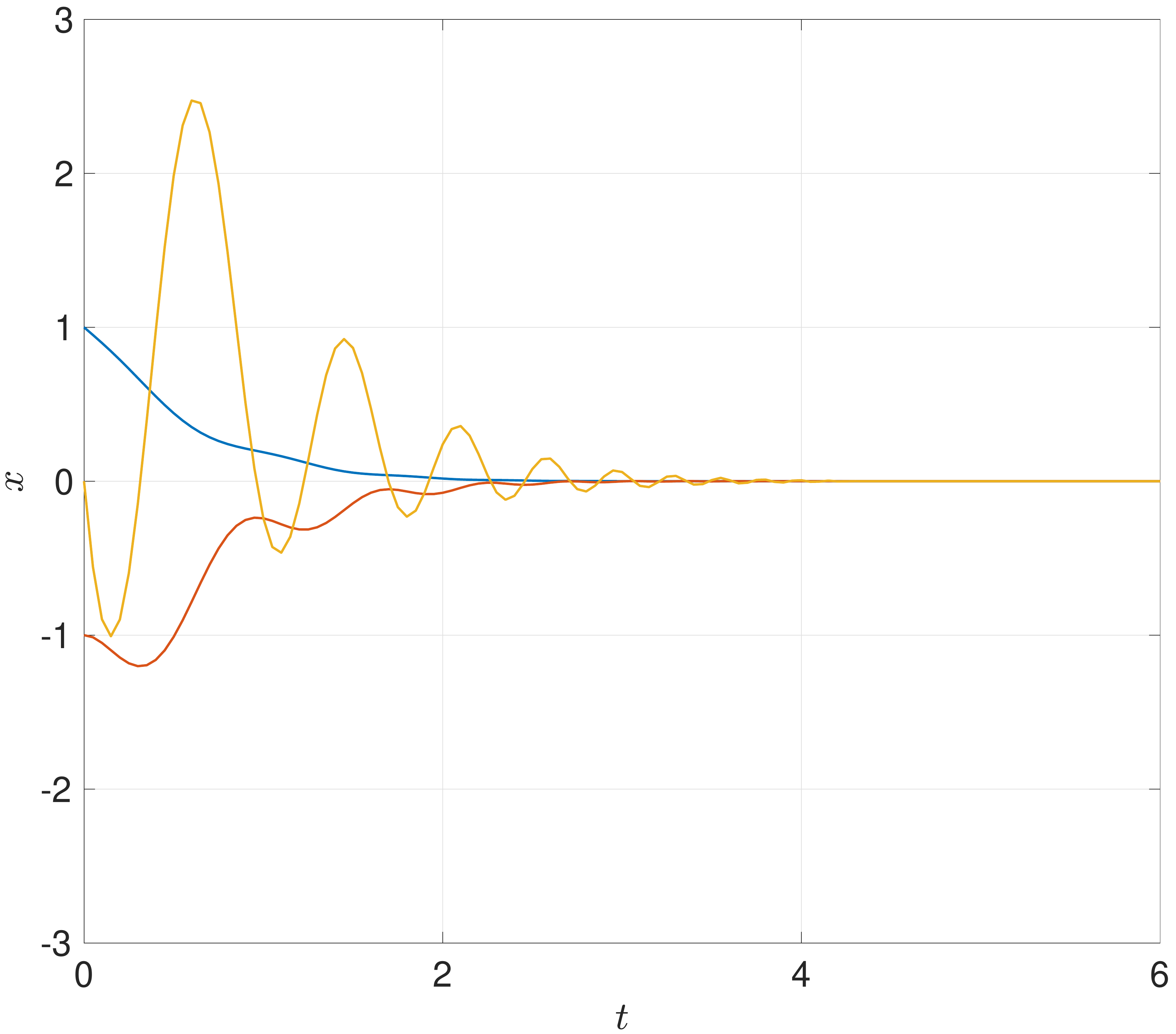}\\
	\includegraphics[width=41mm,height=27mm, bb=0 0 47cm 30cm]{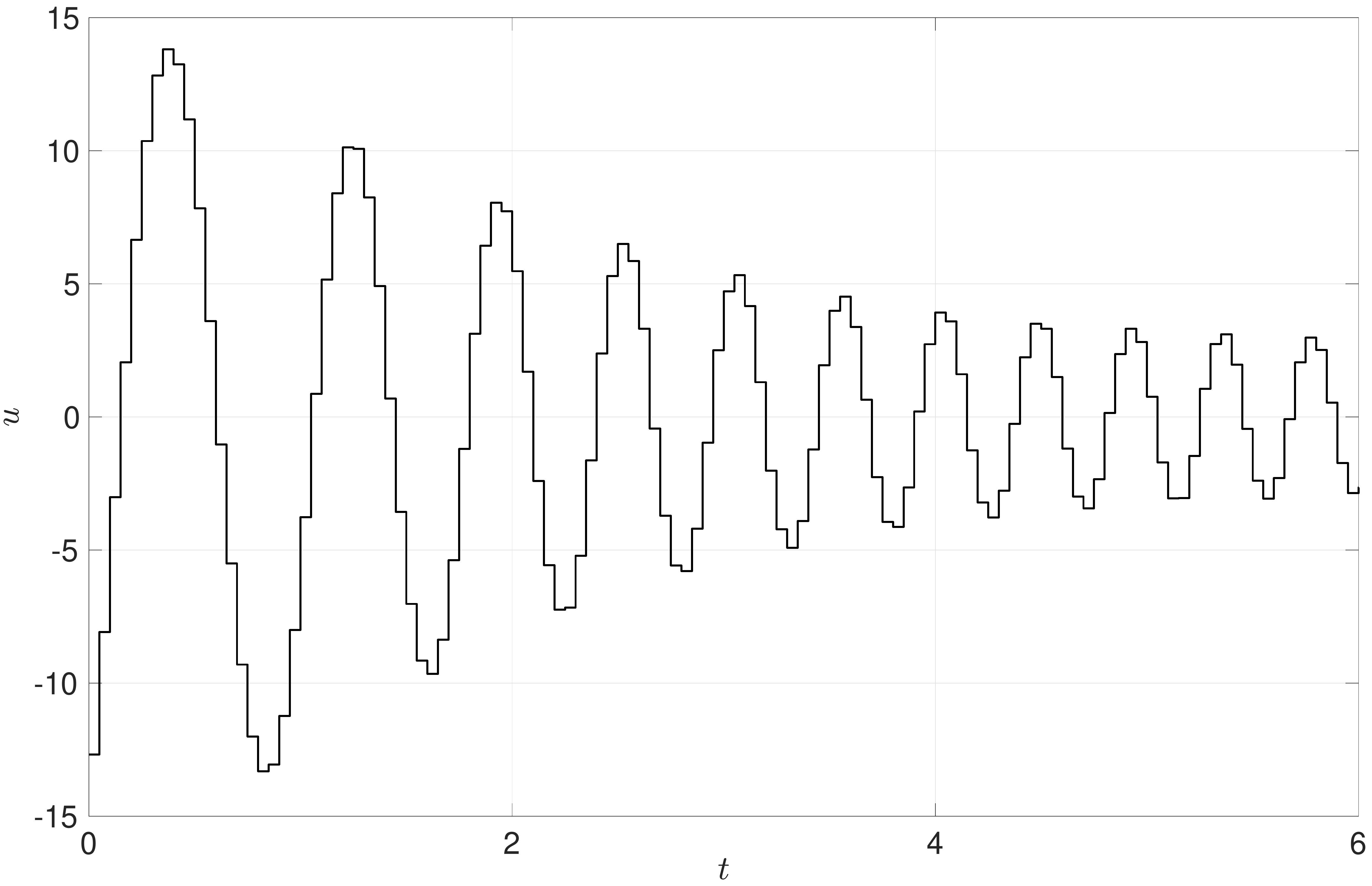}
	\includegraphics[width=41mm,height=27mm, bb=0 0 47cm 30cm]{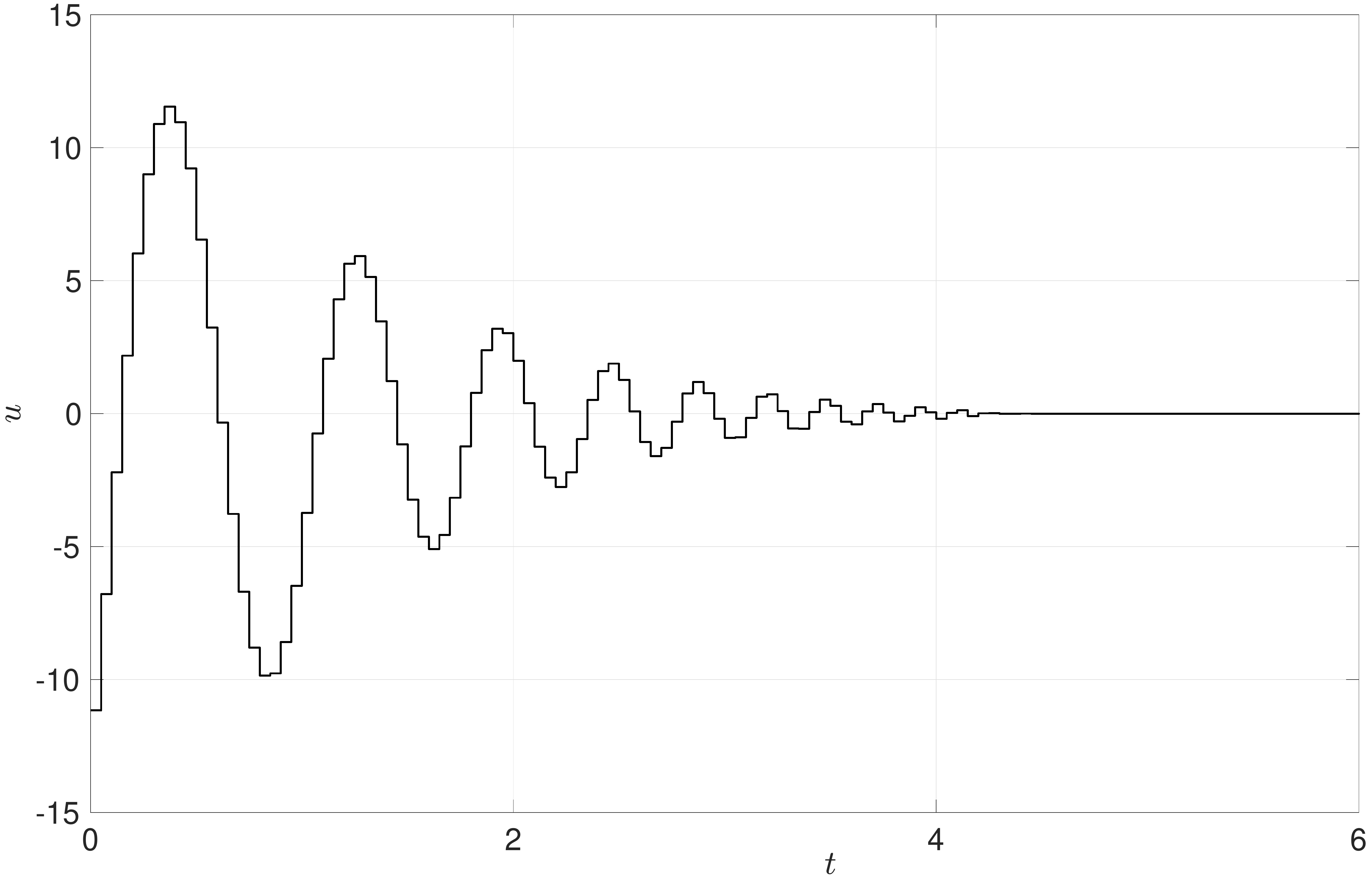}
	\caption{ The simulation results for the system \eqref{eq:main_system} with explicitly (left) and consistently (right) discretized finite-time control \eqref{eq:hom_control} for $n=3,m=1,\mu=-0.25, \rho=1,h=0.05$.}
	\label{fig:n3_no_pert}
\end{figure}

\begin{figure}[h!]
	\centering
	\includegraphics[width=41mm,height=27mm, bb=0 0 40cm 28cm]{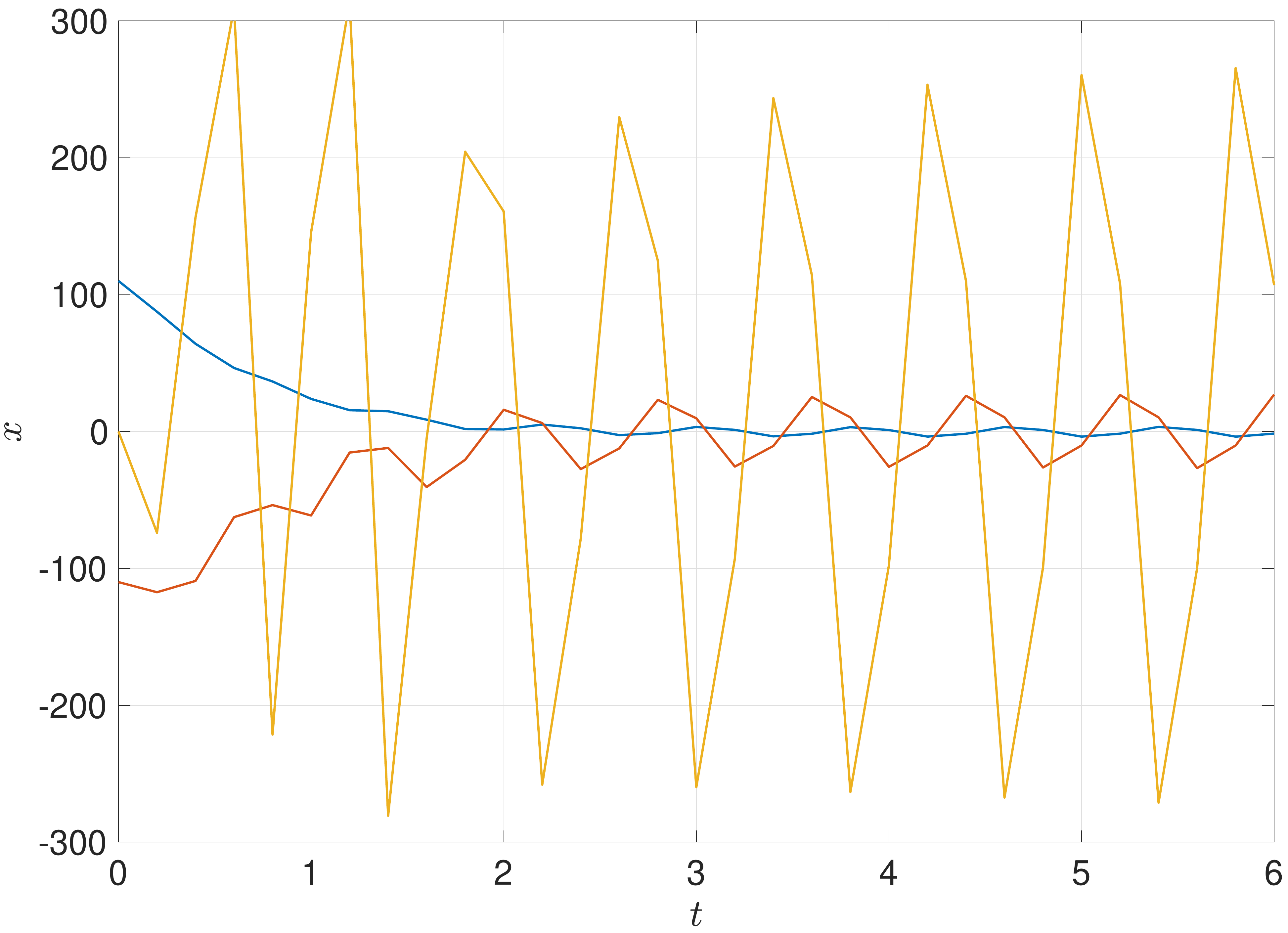}
	\includegraphics[width=41mm,height=27mm, bb=0 0 48cm 30cm]{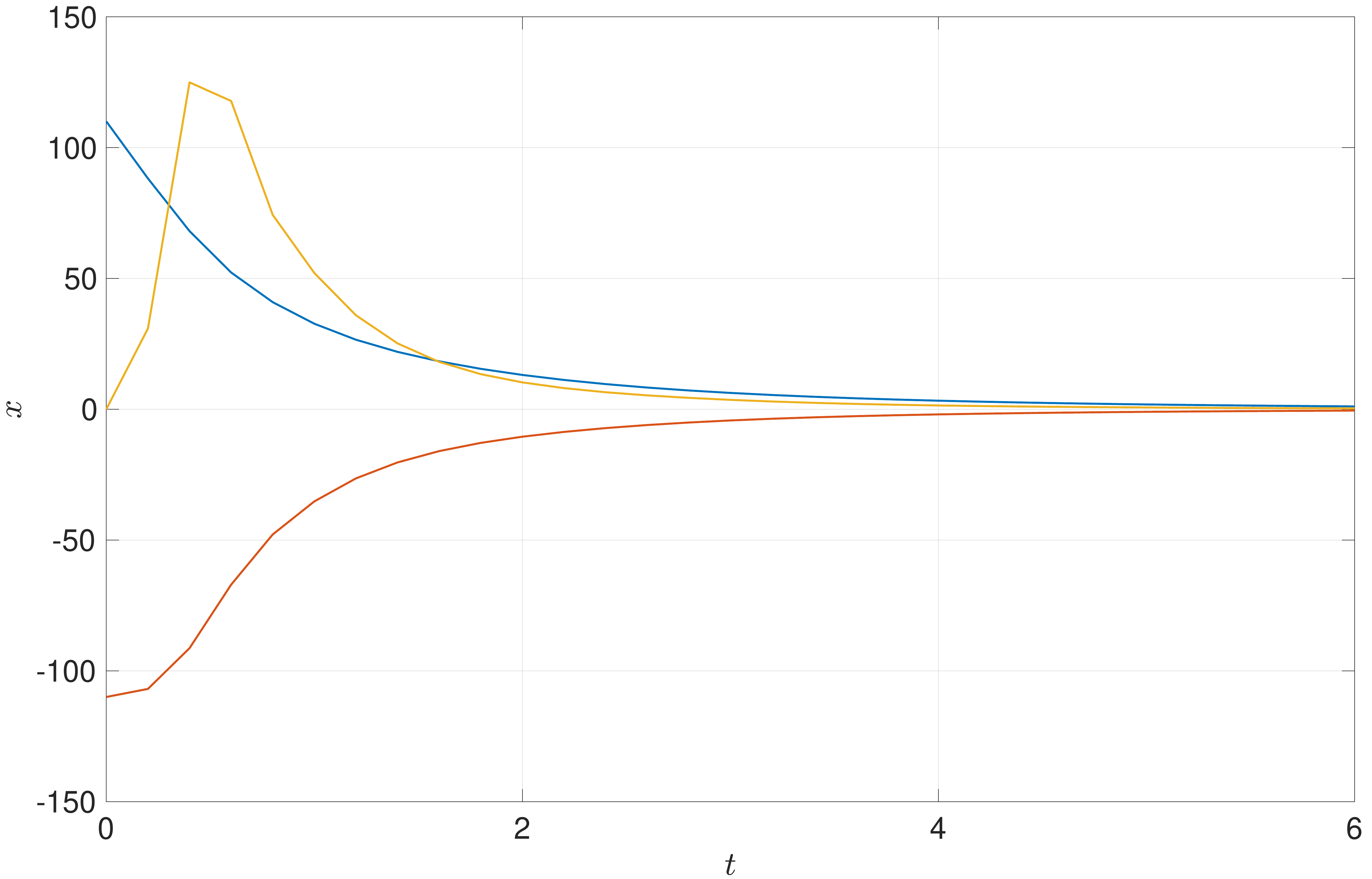}\\
	\includegraphics[width=41mm,height=27mm, bb=0 0 40cm 30cm]{u_exp_n3_fxt.pdf}
	\includegraphics[width=41mm,height=27mm, bb=0 0 47cm 28cm]{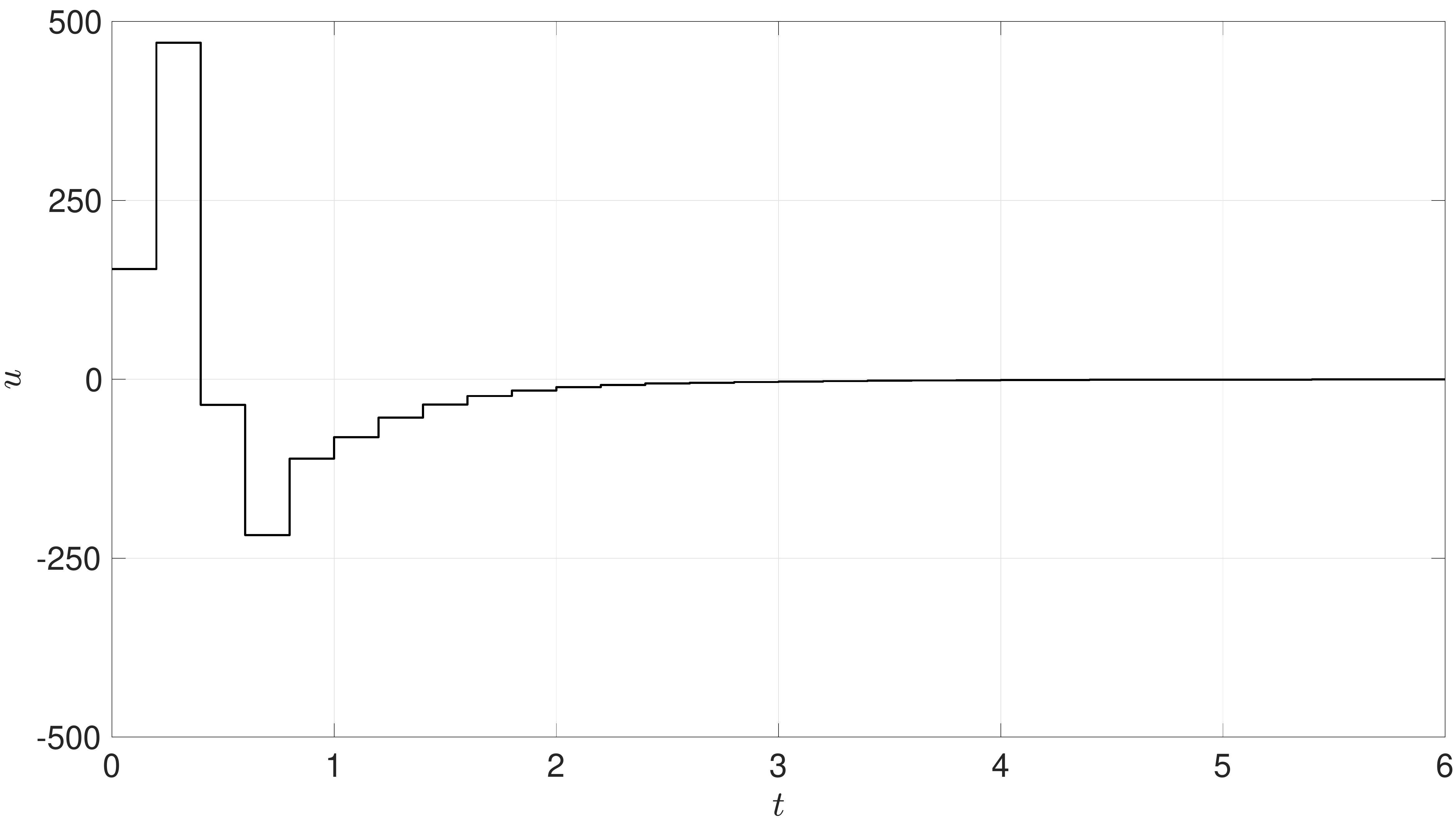}
	\caption{The simulation results for the system \eqref{eq:main_system} with explicitly (left) and consistently (right) discretized fixed-time control \eqref{eq:hom_control} for $n=3,m=1,\mu=0.25, \rho=1,h=0.2$.}
	\label{fig:n3_no_pert_fxt}
\end{figure}

Let $n=3$, $A=\left[ \begin{smallmatrix}   \zero & I_2\\ 0 & \zero \end{smallmatrix}\right]$,  $\mu=-0.25$, $\rho=1$. By Theorem  \ref{thm:hom_control} we derive   a finite-time homogeneous control \eqref{eq:hom_control} with parameters $K_0=\zero$, $G_{\dn}=\left[ \begin{smallmatrix}  1.5 &  & 0\\ 0 & 1.25 &0\\ 0 & 0 & 1\end{smallmatrix}\right]$ and  $X=\left[\begin{smallmatrix}
	1.0000  & -1.5000  &  0.6063\\
	-1.5000  &  3.5187  & -4.3984\\
	0.6063  & -4.3984 &  49.3488\\ 	
\end{smallmatrix}\right]$, $Y=[\begin{smallmatrix}2.8828  & -39.4523 & -49.3488\end{smallmatrix}]$. Simulation results of the system \eqref{eq:main_system} with the consistently discretized control \eqref{eq:sampled_con}, \eqref{eq:hom_consist_reduced_MIMO} for $x(0)=[1\;\; \text{--}1 \;\; 0]^{\top}$, $h=0.05$  are presented on Fig. \ref{fig:n3_no_pert} (right). The system states converges to zero in  a finite time: $x(t)=\zero, u(t)=\zero$ for  $t\geq 4.65$. The results  for the \textit{explicit discretization} 	\vspace{-2mm}
\begin{equation}
	\label{eq:exp_dicr}
	u(t)=\tilde u(x(t_k)), \quad  t\in [t_k,t_{k+1})	\vspace{-2mm}
\end{equation}
of the control \eqref{eq:hom_control} are depicted on Fig. \ref{fig:n3_no_pert} (left)
for comparison reasons. The system \eqref{eq:main_system} with the explicitly discretized homogeneous controller is not asymptotically  stable and suffers of a chattering.
For $\mu=0.25$, $\rho=1$ we similarly derive  a nearly fixed-time homogeneous control \eqref{eq:hom_control} with parameters
$K_0=\zero$, $G_{\dn}=\left[ \begin{smallmatrix}  0.5 &  & 0\\ 0 & 0.75 &1\\ 0 & 0 & 1\end{smallmatrix}\right]$ and  $X=\left[\begin{smallmatrix}
	1.0000  & -0.5000  & -0.8854\\
	-0.5000   & 1.5104  & -1.1328\\
	-0.8854   & -1.1328  &  8.5707\\
\end{smallmatrix}\right]$, $Y=[ \begin{smallmatrix}  2.4609 & -6.5883 &  -8.5707\end{smallmatrix}]$. 
The simulation results for  the explicitly discretized controller with $x(0)=[110\;\; -\!110]^{\top}$ are depicted in Fig. \ref{fig:n3_no_pert_fxt} (left). With this discretization, the system simply blows up for larger initial conditions. Simulations of the consistently discretized control  were made for initial conditions  with various magnitudes up to $\|x_0\|\approx 10^{10}$. They show that the nearly fixed-time stability of the closed-loop system is preserved  as in Fig. \ref{fig:n3_no_pert_fxt} (right).
ISS of both controllers (for $\mu=-0.25$ and $\mu=0.25$) with respect to additive perturbations and measurement noises was also confirmed by simulations. 

\subsection{Multi-input  system}

\begin{figure}[h!]
	\centering
	\includegraphics[width=41mm,height=27mm, bb=0 0 39cm 30cm]{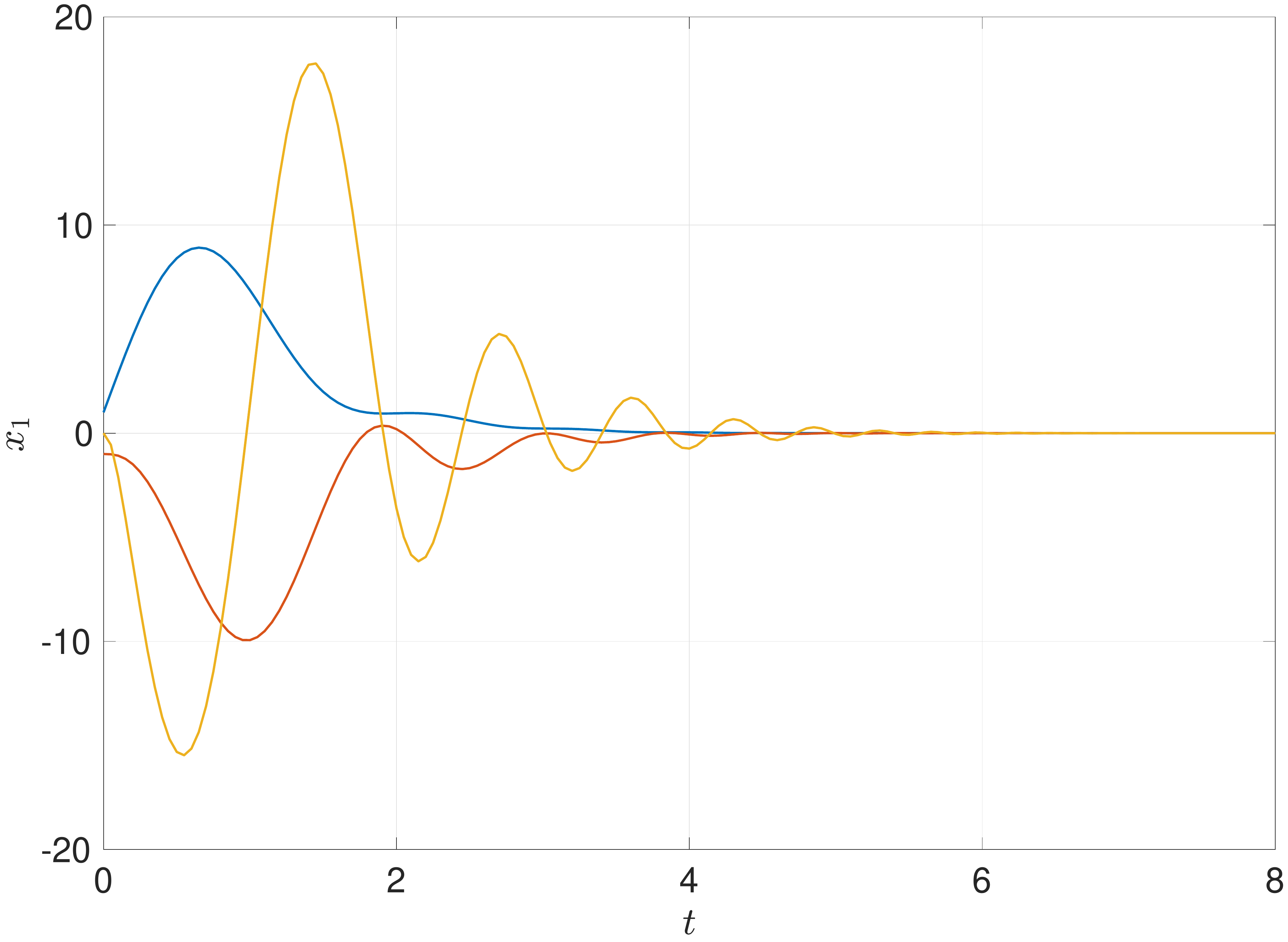}
	\includegraphics[width=41mm,height=27mm, bb=0 0 40cm 30cm]{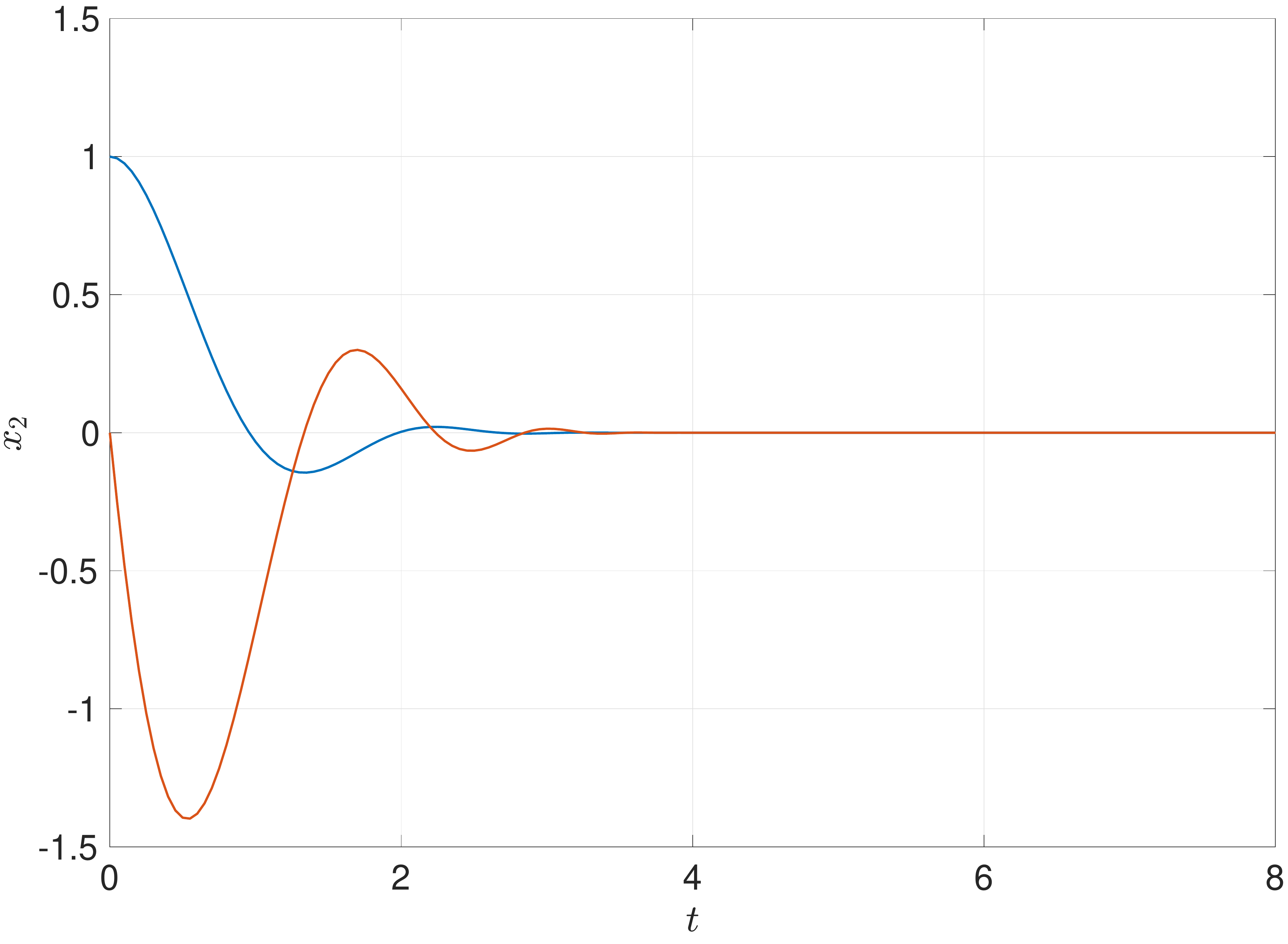}\\
	\includegraphics[width=41mm,height=27mm, bb=0 0 39cm 30cm]{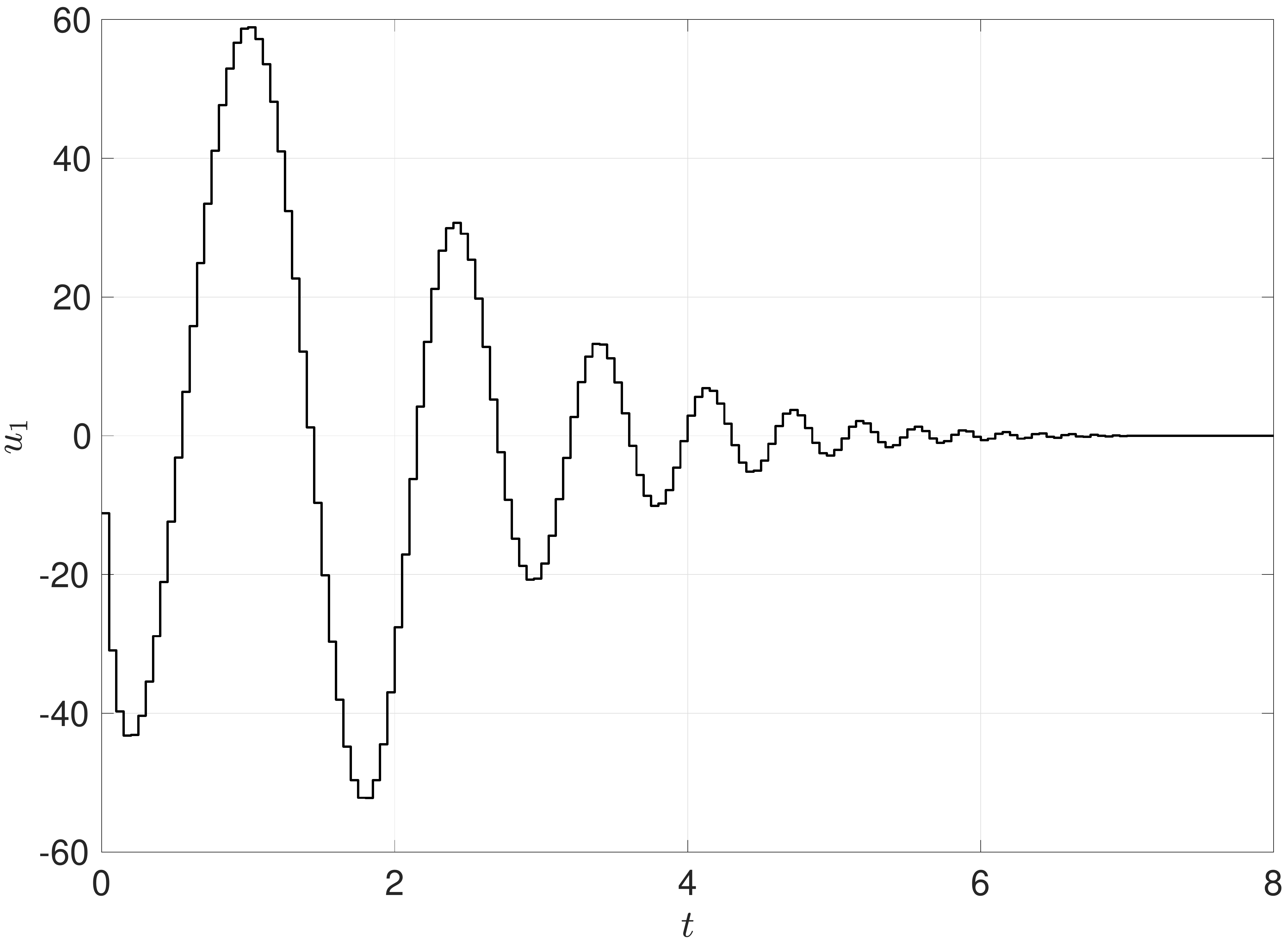}
	\includegraphics[width=41mm,height=27mm, bb=0 0 39cm 30cm]{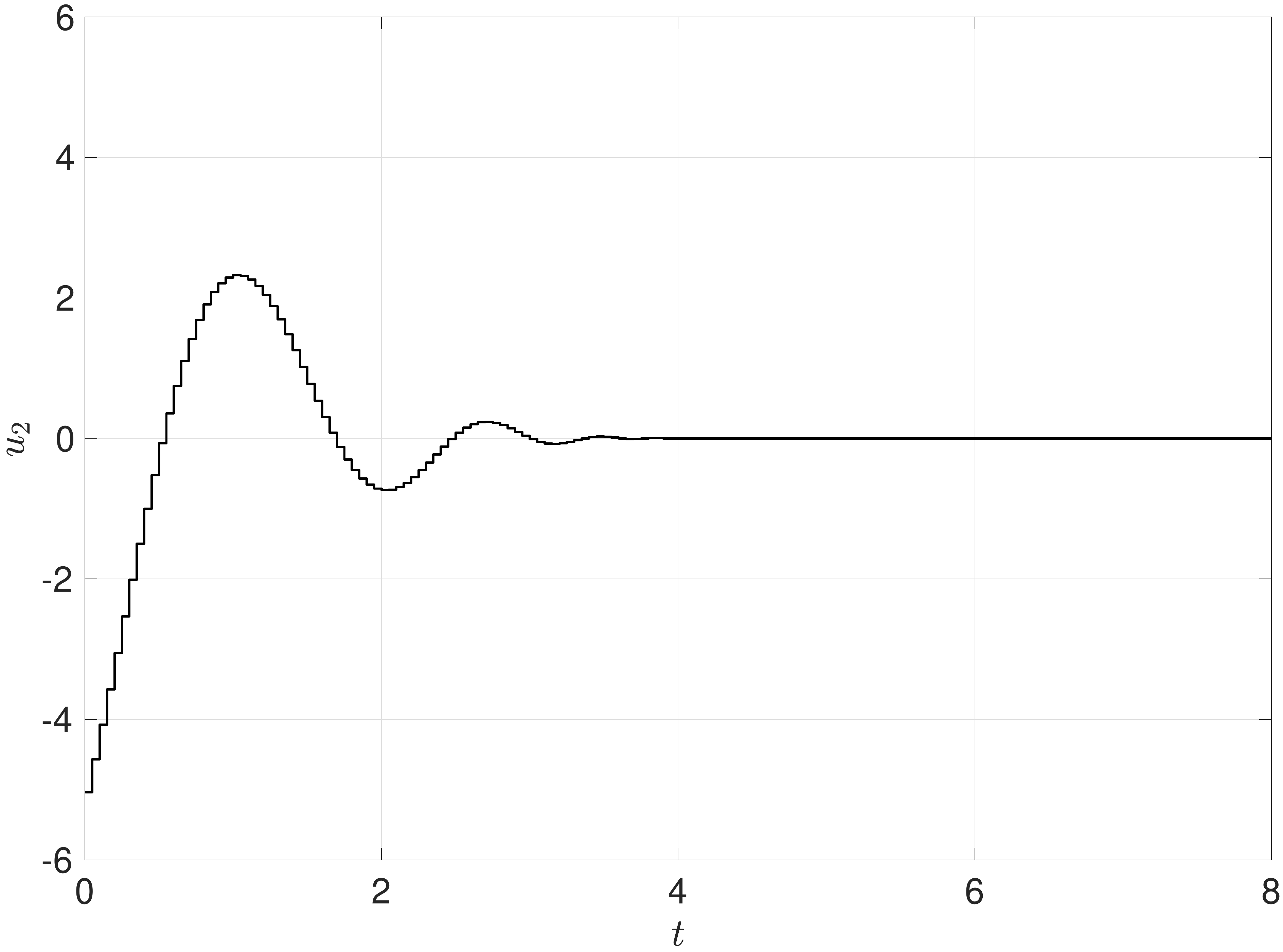}
	\caption{The simulation results for the cascade system \eqref{eq:main_system} with  consistently discretized finite-time controllers with $n_1\!=\!3,n_2\!=\!2,\mu_1\!=\!-0.25, \rho_1\!=\!1,\mu_2\!=\!-1,\rho_2\!=\!-1,h\!=\!0.05$.}
	\label{fig:casc}
\end{figure}

For the multi-input case we consider the above single-input system  ($n_1=3$)  with the finite-time  control ($\mu_1=-0.25$) in the cascade with the second order system ($n_2=2$) with the finite-time control ($\mu_2=-1$)
considered in Remark \ref{rem:feas}, i.e.,
$
A=\left[ \begin{smallmatrix}  
	A_1 & A_{12}\\
	\zero & A_2 
\end{smallmatrix}\right]\!, B=\left[ \begin{smallmatrix}  B_1  & \zero \\ \zero  & B_2\end{smallmatrix}\right]\!,
A_1=\left[ \begin{smallmatrix}   \zero & I_2\\ 0 & \zero \end{smallmatrix}\right],	A_{12}=\left[
\begin{smallmatrix}
	1 & 0 \\ 0 & 0 \\ 0 & 0 
\end{smallmatrix}\right]\!,
$
$A_2=\left[ \begin{smallmatrix}   0 &1\\ 0 & \zero \end{smallmatrix}\right], B_1=\left[ \begin{smallmatrix}   0 \\ 0\\ 1 \end{smallmatrix}\right], B_2=\left[ \begin{smallmatrix}   0 \\ 1 \end{smallmatrix}\right]. $ The simulation results for the cascade system are shown on Fig. \ref{fig:casc} for $x_1(0)=[1\;\; -\!1\;\; 0]^{\top}, x_2(0)=[1 \;\; 0]^{\top}$. They confirm finite-time stability of the closed-loop system with the consistently discretized  control: $x_1(t)=\zero,u_1(t)\!=\!0,\forall t\!\geq\! 7.35$ and $x_2(t)\!=\!\zero, u_2(t)\!=\!0,\forall t\!\geq\! 4.2$.   
\vspace{-2mm}

\section{Conclusions} 
\label{sec:6}
In the paper, two types of discretization (sampled-time implementation) schemes for a homogeneous control law are developed.  Both preserves the finite/fixed-time stability properties of the original continuous-time  closed-loop system.
The first scheme gives a mixture of a feedforward and a feedback algorithms. It can be utilized if the control sampling is much faster than the  sampling of measurements, {or in the model-predictive framework}.  The second scheme (called consistent) provides a static feedback law, which always preserves the stability properties, at least, locally  (close to zero and close to infinity).
A necessary and sufficient condition of global consistency is presented. A particular sufficient condition is formalized in terms of a parametric LMI, which, by numerical simulations, is shown to be feasible in some cases. 
A development of numerical algorithms for control parameters tuning based on the obtained conditions of the consistency is an interesting problem for the future research.

\section{Appendix}
\subsection{Auxiliary results}
\begin{lemma}\cite[page 136]{Hall2015:Book}\label{lem:exp_XY}
	If $Z_1Z_2-Z_2Z_1=qZ_2$ with $q\in \R$ and $Z_1,Z_2\in \R^{n\times n}$ then 
	$
	e^{Z_1}e^{Z_2}=e^{Z_1+\frac{q}{1-e^{-q}}Z_2}.
	$
\end{lemma}
\begin{lemma}\label{lem:main}
	If Assumption \ref{as:1} is fulfilled and $\dn$  is a linear dilation defined in Theorem \ref{thm:hom_control} then \vspace{-2mm}
	\begin{itemize}		
		\item[1)] for all $s\in \R$ the following identities hold:\vspace{-2mm}
		\begin{equation}\label{eq:d_A_B}
			\begin{smallmatrix}\!\!\!\!\!\!A\dn(s)=e^{\mu s} \dn(s) A, & \;\;\dn(s)B=e^sB, &\;\;	\dn(s)e^{\tau A}=e^{e^{-\mu s}\tau A}\dn(s),
			\end{smallmatrix}\!\!\!\vspace{-2mm}
		\end{equation}	
		\begin{equation}\label{eq:int_e}
			\begin{smallmatrix}
				\dn(s)\int^h_0 e^{\tau A}d\tau \,=\,e^{\mu s}\int^{e^{-\mu s}h}_0 e^{\tau A} d\tau  \dn(s);
			\end{smallmatrix}
		\end{equation}
		\item[2)] the matrix $W_{h}$ given by \eqref{eq:W}  satisfies	\vspace{-2mm}
		\begin{equation}\label{eq:Wh}
			\begin{smallmatrix}
				W_h=\dn_*(\ln h) \int^1_0e^{\tau A}d\tau  \left[B\;\; e^{A}B\;...\; e^{(n-1)A}B\right],
			\end{smallmatrix}
			\vspace{-2mm}
		\end{equation}
		\begin{equation}\label{eq:W_inv}
			\begin{smallmatrix}W_h^{-1}\dn_*(\ln h)\int^{n}_0 e^{\tau A} d\tau B\end{smallmatrix}=\left[
			\begin{smallmatrix}
				1\\			
				...\\
				1\end{smallmatrix}
			\right],
		\end{equation}   
		where $\dn_*$ corresponds to $\dn$ with $\mu=-1$;
		\item the parameter $\beta$ defined in Theorem \ref{thm:hom_norm} satisfy $\beta=1$  for any $\mu<0$.
	\end{itemize}
\end{lemma}
\textbf{Proof.}
\textbf{1)} Under Assumption \ref{as:1}, Theorem \ref{thm:hom_control} the dilation $\dn(s)=e^{sG_{\dn}}$ is uniquely defined such that   $G_{\dn}$ satisfies \eqref{eq:hom_A0} with $A_0=A$. We derive
$
A\dn(s)\!=\!A\sum_{i=0}^{+\infty} \tfrac{s^iG_{\dn}^i}{i!}\!=\!\sum_{i=0}^{+\infty} \tfrac{s^i(\mu I_n+G_{\dn})^iA}{i!}\!=\!e^{\mu s}\dn(s)A,
$
$
\dn(s)B=\!\sum\limits_{i=0}^{+\infty} \tfrac{s^iG_{\dn}^i}{i!}B=\!\sum\limits_{i=0}^{+\infty} \tfrac{s^i}{i!}B, $ $\forall s\in \R.
$
Hence, for all $s\in \R$ we have
$
\dn(s) e^{\tau A} \!=\! \dn(s)\!\sum\limits_{i=0}^{n-1}\! \tfrac{h^{i}A^i}{i!}\!=\!\sum\limits_{i=0}^{n-1} \!\tfrac{e^{-\mu is} h^{i}A^i \dn(s)}{i!}\!=\!e^{e^{-\mu s}\tau A}\dn(s)
$
and 
$
\dn(s)\int^h_0 e^{\tau A}d\tau =\int^h_0 e^{e^{-\mu s}\tau A}\dn(s) d\tau.
$
and changing of the integration variable we derive  \eqref{eq:int_e}.
\textbf{2)} On the one hand, since $e^{sA}$ and $\int^h_0 e^{\tau A}d\tau$ commute for any $s\in \R$ then taking into account $A_h^i=e^{ihA}$ we derive 
$
W_h=[B_h\;\; A_hB\;\; A_h^2B_h\;\; ...\;\;A_h^{n-1}B_h]=
$
$	
\left(\int^h_0 e^{sA}ds \right)\cdot$ $\cdot [B;\; e^{hA}B\;\; e^{2hA}B \;\;...\;\; e^{(n-1)hA}B].
$
On the other hand, the homogeneity identities \eqref{eq:d_A_B} imply $hB=\dn_*(\ln h)B$ and 
$e^{ihA}\dn_*(\ln h)=\dn_*(\ln h)e^{iA}, \forall i\in \N, \forall h>0$. Hence, we have\\
$
W_h\!=\!\left(\!\frac{1}{h}\!\int^h_0 \!\!e^{sA}ds \!\right)\!\dn_*(\ln h)[B,e^{A}B,e^{2A}B,...,e^{(n-1)A}B].
$
%	\[
%	\left(\frac{1}{h}\!\int^h_0 \!\!e^{sA}ds \right)\dn_*(\ln h) \sum_{i=0}^{n-1} e^{iA}Be_{i+1}^{\top},
%	\]
Using the identity \eqref{eq:int_e} we derive\\
\centerline{$
	\dn_*(-\ln h)   \left(\frac{1}{h}\!\int^h_0 \!\!e^{sA}ds \right)=\int^1_0 \!\!e^{sA}ds \;\dn_*(-\ln h),
	$}
so the formula \eqref{eq:Wh} holds.	Finally, since 
{\scriptsize $
	\int^{n}_0 \!\!e^{\tau nA} d\tau =\sum_{i=0}^{n-1} \int_i^{i+1} \!\!e^{\tau A} d\tau\!=\!\!
	\int_0^1 \!e^{\tau A} d\tau\sum\limits_{i=0}^{n-1} \!e^{iA}
	$}
then
{\scriptsize $
	W_h^{-1} \dn_*\!(\ln h)\cdot $ $\cdot \int^{n}_0 \!e^{\tau n A} d\tau B\!=\!
	[B\,e^A\!B\,...\,e^{(n-1)A}\!B]^{\text{--}1}\!\sum_{i=0}^{n\text{--}1} \!e^{iA}\!B\!=\!\sum_{i}^n \!e_i.
	$}\\
\textbf{3)} For $m=1$, under conditions of  Theorem \ref{thm:hom_control}, we have $G_{\dn}=U^{-1} \diag\{1-\mu(n-1), ..., 1\}U$ for some non-singular matrix $U\in \R^{n\times n}$. The latter means that $\beta=1$ 
\begin{lemma}\label{lem:V}
	The function $V:\R^n\mapsto [0,+\infty)$ defined as 
	$V(x)=\overline{\sigma}^{-1}(\|x\|_{\dn}), x\in \R^n$ is positive definite radially unbounded and globally Lipschitz continuous with the Lipschitz constant $1$,
	where $\overline{\sigma}\in\mathcal{K}_{\infty}$ is as in Theorem \ref{thm:hom_norm}.
\end{lemma}
\textbf{Proof.} The Lipschitz continuity for $\|x\|\geq 1$ follows from \cite[Proposition 1]{Polyakov2018:RNC}.
For $\|x\|\leq 1$ this fact can be proven similarly (just replacing $\beta$ with $\alpha$). All other properties of $V$ are obvious.
\begin{lemma} \label{lem:u_u_h}  For any $C_*\!>\!0$ and any $\varepsilon\in (0,1)$ there exist $C>\!0$ and $r\!>\!0$ such that \vspace{-2mm}
	\begin{equation}\label{eq:u_u_h}
		\|u(x)-u_{\hat h}(\tilde x)\|\leq C \min\{\|x\|_{\dn}^{1+2\mu}, \|\tilde x\|^{1+2\mu}\},\vspace{-2mm}
	\end{equation}
	for all $x,\tilde x$ satisfying  $(1-\varepsilon)\|\tilde x\|_{\dn}\!\leq\! \|x\|_{\dn}\!\leq\! (1+\varepsilon) \|\tilde x\|_{\dn}$,  $\|x\|^{\mu}_{\dn}\!\leq\! r$, $ \|\dn(-\ln \| x\|_{\dn})(x-\tilde x)\|\!\leq\! C_*\| x\|^{\mu}$.
\end{lemma}
\textbf{Proof.} 
Using approximation of $\tilde u_{\tilde h}$ (see, the proof of Proposition \ref{prop:approx}) we conclude $\exists h_0\in (0,1), \exists C_0>0$ such that $\sup_{y\in S} |u(y)-\tilde u_{\tilde h}(y)|\leq C_0\tilde h$, $\forall  \tilde h<h_0$. 
Since $u$ is locally Lipschitz continuous on $\R^{n}\backslash\{\zero\}$ then 
for any $\delta\in (0,1)$ there exists $C_1$ such that $|u(y)-u(z)|\leq C_1\|y-z\|$,  $\forall y\in S$ and $\forall z: \|y-z\|\leq\delta$. Using the Lipschitz condition and  the dilation symmetry of $u$ and $\tilde u_h$
(see \eqref{eq:hom_tilde_u}) we derive \vspace{-2mm}
\[
\begin{smallmatrix}
	|u(x) -\tilde u_{\hat h}(\tilde x)|\leq |u(x)-u(\tilde x)|+|u(\tilde x) -\tilde u_{\hat h}(\tilde x)|=\\
	\|x\|^{1+\mu}_{\dn}|u(\dn(-\ln \|x\|_{\dn})x)-u(\dn(-\ln \|x\|_{\dn})\tilde x)|+ \\
	\|\tilde x\|_{\dn}^{1+\mu} |u(\dn(-\ln \|\tilde x\|_{\dn})\tilde x)- u_{\|\tilde x\|_{\dn}^{\mu}\hat h}(\dn(-\ln \|\tilde x\|_{\dn})\tilde x)|\leq \\
	C_1  \|x\|^{1+\mu}_{\dn} \|\dn(\text{--}\ln \|x\|_{\dn})x-\dn(\text{--}\ln \|x\|_{\dn})\tilde x\|+ \|\tilde x\|_{\dn}^{1+\mu}C_0\|\tilde x\|_{\dn}^{\mu}\hat h=\\
	C_1  \|x\|^{1+\mu}_{\dn} \|\dn(\text{--}\ln \| x\|_{\dn})x-\dn(\text{--}\ln \| x\|_{\dn})\tilde x\|+C_2 \|\tilde x\|_{\dn}^{1+2\mu}
\end{smallmatrix}\vspace{-2mm}
\]
provided that $\|\dn(-\ln \|x\|_{\dn})x-\dn(-\ln \|x\|_{\dn})\tilde x\|\leq \delta$ and $\|\tilde x\|_{\dn}^{\mu}\leq h_0/\hat h$, where $C_2=C_0\hat h$. Taking $r\!\leq\!  \left(\frac{1-\varepsilon}{1+\varepsilon}\right)^{\!|\mu|}\!\!\min\!\left\{\frac{\delta}{C_*}, \frac{h_0}{\hat h}\right\}$ and $C\!=\!(C_1C_*+C_2)\left(\frac{1+\varepsilon}{1-\varepsilon}\right)^{1+2|\mu|}$ we complete the proof.
\subsection{The proof of Corollary \ref{cor:hom_solution}}
Denoting 
$
y=\|x\|\dn(-\|x\|_{\dn})x,
$
we derive
$
\|y\|=\|x\|_{\dn}\|\dn(-\ln \|x\|_{\dn})x\|=\|x\|_{\dn}
$
and conclude that the closed-loop system \eqref{eq:main_system}, \eqref{eq:hom_control} is
topologically equivalent (homeomorphic on $\R^n$ and diffeomorphic on $\R^{n}\backslash\{\zero\}$, see \cite{Polyakov2018:RNC}) to the standard homogeneous system:\vspace{-2mm}
\[
\dot y=
\|y\|^{\mu} (A_0+BK+\rho(G_{\dn}-I_n)) y,\vspace{-2mm}
\]
where the identities  
$\dn(s)A_0=e^{-\mu s}A_0\dn(s)$ and $\dn(s)B=e^{s}B$, $\forall s\in \R$ 
are utilized on the last step.
In this case, using \eqref{eq:LMI} we conclude\vspace{-2mm}
\[\tfrac{d}{dt} \|y(t)\|\!=
%\!\tfrac{y^\top\! P \dot y(t)}{\|y(t)\|}
%	\!=\!\|y\|^{\mu-1}
%	y^\top\!P\left(A_0\!+\!BK\!+\!\rho(G_d-I_n)\right)\!y
%	\]
%	\[
%	=	\!\|y\|^{\mu-1}\!\!
%	\left(\!\tfrac{y^\top\!\left\{\!P\left(A_0+BK+\rho G_{\dn}\right) +\left(A_0+BK+\rho G_{\dn}\right)^{\top}P\!\right\} y}{2} \text{--}\rho y^{\!\top}\!\!Py\!\right)
%	\]
%	\[=	
-\rho \|y(t)\|^{\mu+1}\vspace{-2mm}
\]
and
$
\|y(t+\tau)\|^{-\mu}=\|y(t)\|^{-\mu}+\mu\rho\tau,
$
for $\|y(t)\|^{-\mu}+\mu\rho\tau\geq 0$. Obviously, $\|y(t+\tau)\|=0$ if
$\|y(t)\|^{-\mu}+ \mu\rho \tau\leq 0$. The latter corresponds to the negative homogeneity degree $\mu<0$ and the finite-time stability of the closed-loop system. Hence, denoting $\tilde A=A_0+BK+\rho G_{\dn}$  
we obtain\vspace{-2mm}
\[
\begin{array}{c}
	y(t+\tau)=e^{(\tilde A-\rho I_n))\int^{\tau}_{0} \|y(t+\sigma)\|^{\mu} d\sigma}y(t)\\
	%=&e^{(A+BK_{lin}+\rho(G_{\dn}-I_n))\int^{\tau}_{0} \frac{1}{\|y(t)\|^{-\mu}+\mu \rho \sigma} d\sigma}y(t)\\
	%=&e^{(A+BK_{lin}+\rho(G_{\dn}-I_n))\frac{1}{\mu\rho }\ln \frac{\|y(t)\|^{-\mu}+\mu\rho \tau}{\|y(t)\|^{-\mu}}}y(t)\\
	=e^{(\tilde A-\rho I_n)\frac{1}{\mu\rho}\ln (1+\mu\rho\tau\|y(t)\|^{\mu})}y(t).
\end{array}\vspace{-2mm}
\]
Since $	\|y(t)\|=\|x(t)\|_{\dn}$ then returning to the original coordinates  we derive    \vspace{-2mm}     
\[
%	\begin{split}
x(t+\tau)=
\dn (\ln \|y(t+\tau\|)\tfrac{y(t+\tau)}{\|y(t+\tau)\|}=
%	=&
%	\dn(\ln(\|y(t)\|^{-\mu}+\mu \rho
%		\tau)^{\frac{1}{-\mu}}) \\
%		&\cdot e^{(A+BK_{lin}+\rho(G_{\dn}-I_n))\frac{1}{\rho \mu}\ln (1+\mu\rho \tau\|y(t)\|^{\mu})}\\
%		&\cdot
%		\tfrac{y(t)}{(\|y(t)\|^{-\mu}+\mu\rho \tau)^{\frac{1}{-\mu}}}\\
%		=&	\dn(\ln(\|x(t)\|_{\dn}^{-\mu}+\mu
%		\rho\tau)^{\frac{1}{-\mu}})\\
%		&\cdot e^{(A+BK_{lin}+\rho G_{\dn})\frac{1}{\rho \mu}\ln (1+\mu\rho\tau\|x(t)\|_{\dn}^{\mu})}\\
%		&\cdot 
%		\dn(-\ln \|x(t)\|_{\dn})x(t)\\
%		=&
Q_{\tau}(\|x(t)\|_{\dn})x(t)\vspace{-2mm}
%	\end{split}
\]
for all $t\geq 0$ and all $\tau\geq 0$.

\subsection{The proof of Corollary \ref{cor:cascade}}
Under Assumption \ref{as:2}, the model \eqref{eq:main_system} is a system of interconnected systems
with state vectors $x_i\in \R^{n_i}$	
and control inputs $u_i\!\in\! \R$.
By the formula \eqref{eq:hom_norm_deriv}, we derive	\vspace{-2mm}
\[
\tfrac{d}{dt}\|x_i\|_{\dn_i}=-\rho_i \|x_i\|_{\dn_i}^{1+\mu_i}+	\vspace{-2mm}
\]
\[
\|x_i\|_{\dn_i}\tfrac{x^{\top}_i\dn^{\top}_i(-\ln \|x_i\|_{\dn_i})P_i\dn_i(-\ln \|x_i\|_{\dn_i})\sum_{j=i+1}^{m} A_{ij}x_j}{x^{\top}_i\dn^{\top}_i(-\ln \|x_i\|_{\dn_i})P_iG_{\dn_i}\dn_i(-\ln \|x_i\|_{\dn_i})x_i},
\]
where $A_{ij}$ are over diagonal blocks of the matrix $A$.  By Theorem \ref{thm:hom_control} the norm in $\R^{n_i}$ is defined as $\|z\|=\sqrt{z^{\top}P_iz}$, $z\in \R^{n_i}$ and
% by definition of the canonical homogeneous norm, we have  $x_i^{\top}\dn^{\top}_i(-\|x_i\|_{\dn_i})P_i\dn_i(-\|x_i\|_{\dn_i})x_i=1$, or equivalently, 
$\|\dn_i(-\|x_i\|_{\dn_i})x_i\|=1$. Taking into account $P_iG_{\dn_i}+G_{\dn_i}^{\top}P_i\succ 0$ we derive 	\vspace{-2mm}
\[
x^{\top}_i\dn^{\top}_i(-\ln \|x_i\|_{\dn_i})P_iG_{\dn_i}\dn_i(-\ln \|x_i\|_{\dn_i})x_i\geq\beta_i>0,	\vspace{-2mm}
\]
where $\beta_i=0.5\lambda_{\min} (P_i^{1/2}G_{\dn_i}P^{-1/2}+P_i^{-1/2}G_{\dn_i}^{\top}P^{1/2}).$
Applying the Cauchy-Schwartz inequality we obtain\vspace{-2mm}
\[\tfrac{x^{\top}_i\dn^{\top}_i(-\ln \|x_i\|_{\dn_i})P_i\dn_i(-\ln \|x_i\|_{\dn_i})\sum_{j=i+1}^{m} A_{ij}x_j}{x^{\top}_i\dn^{\top}_i(-\ln \|x_i\|_{\dn_i})P_iG_{\dn_i}\dn_i(-\ln \|x_i\|_{\dn_i})x_i}\leq\vspace{-2mm}
\]
\[ 
\tfrac{1}{\beta_i}\left\|\dn_i(-\ln \|x_i\|_{\dn_i})\Sigma_{j=i+1}^{m} A_{ij}x_j\right\|\!\leq\! \tfrac{\left\|\Sigma_{j=i+1}^{m} A_{ij}x_j\right\|}{\beta_i\sigma_i(\|x_i\|_{\dn_i})},
\]
where	$\sigma_i=\overline{\sigma}^{-1}$ and $\overline{\sigma}$ is given in Theorem \ref{thm:hom_norm}	
with $\alpha\!=\!\alpha_i\!=\!0.5\lambda_{\max} (P_i^{\frac{1}{2}}G_{\dn_i}P^{-\frac{1}{2}}+P_i^{-1/2}G_{\dn_i}^{\top}P^{1/2})$
and $\beta=\beta_i$.
The estimate obtained for $\frac{d}{dt}\|x_i\|_{\dn_i}$, the stability of the $m$-subsystem and the cascade structure imply the forward completeness of the whole system.

If $V_i(x_i)=\sigma_i(\|x_i\|_{\dn_i})$ then  for $\|x_i\|_{\dn_i}> 1$ we have \\
\centerline{$
	\tfrac{d V_i}{dt} =\beta_i \tfrac{V_i \cdot \frac{d}{dt} \|x_i\|_{\dn}}{\|x_i\|_{\dn}}  \leq -\beta_i \rho_i 
	V^{1+\frac{\mu_i}{\beta_i}}_i+\left\|\sum_{j=i+1}^{m} A_{ij}x_j\right\|.
	$}
For $0<\|x_i\|_{\dn}<1$ we derive\\
\centerline{$
	\tfrac{dV_i}{dt}\!=\!\alpha_i \tfrac{V_i \cdot \frac{d}{dt} \|x_i\|_{\dn}}{\|x_i\|_{\dn}}  \!\leq\! -\alpha_i \rho_i 
	V^{1+\frac{\mu_i}{\alpha_i}}_i +\tfrac{\alpha_i}{\beta_i} \left\|\sum_{j=i+1}^{m} A_{ij}x_j\right\|\!
	$} 
Since $V_i$ locally Lipschitz continuous on $\R^{n_i}\backslash\{\zero\}$ then using the Clarke's gradient  for $\|x_i\|_{\dn_i}=1$ we have \vspace{-2mm}
\[
\begin{split}
	\tfrac{dV_i}{dt}  \leq & -\lambda_i(x_i)\alpha_i \rho_i 
	V^{1+\frac{\mu_i}{\alpha_i}}_i -(1-\lambda_i(x_i))\beta_i \rho_i 
	V^{1+\frac{\mu_i}{\beta_i}}_i\\
	&+\left(\lambda_i(x_i) \tfrac{\alpha_i}{\beta_i}+1-\lambda_i(x_i) \right)\left\|\Sigma_{j=i+1}^{m} A_{ij}x_j\right\|
	\vspace{-2mm}
\end{split}
\] with some $\lambda(x_i)\in [0,1]$.
The latter means that $x_i\mapsto V_i(x_i)$ is an ISS Lyapunov function \cite{SontagWang1996:SCL} of the $i$-th subsystem with respect to the input $\sum_{j=i+1}^{m} A_{ij}x_j$ provided that $\mu_i>-\beta_i$. 
% An asymptotic stability of subsystems with the numbers from $i+1$ to $m$ imply asymptotic stability of the $i$-th subsystem. 
Since the $m$-th subsystem is globally uniformly asymptotically stable then using the cascade structure and  the ISS property of each sysbsystem we conclude that 
the whole system is globally uniformly asymptotically stable as well.    
Moreover, the obtained estimate of $\frac{d}{dt}\|x_i\|_{\dn_i}$ implies that the finite-time, exponential and nearly fixed-time  convergence rates are preserved as well dependently of the sign of $\mu_i$ for all $i=1,...,m$. 	
If $\mu_i\leq -\beta_i$, then the each subsystem with the zero input  is finite-time stable. Taking into account forward completeness and the cascade structure we conclude that the whole system is finite-time stable too.

\subsection{The proof of Theorem \ref{thm:con_full}}

On the one hand, since $x_{k+n}=Q_{nh}(\|x_k\|_{\dn})x_k$ then, in the view of Corollary \ref{cor:hom_solution}, the states of the discrete-time system \eqref{eq:dicrete_model} with the control \eqref{eq:hom_consist_full} coincides with the states of the original continuous-time system \eqref{eq:main_system}, \eqref{eq:hom_control} at time instances $t_{kn}, k=0,1,....$. In this case, if the discrete-time system is globally Lyapunov stable then the finite-time or nearly fixed-time stability property of the original continuous-time system is preserved.

On the one hand, in the view of Theorem \ref{thm:hom_control} and Corollary \ref{cor:hom_solution}, we have   $\|Q_{\tau}(\|x\|_{\dn})x\|^{-\mu}_{\dn}=\|x\|_{\dn}^{-\mu}+\mu\rho \tau$ for all $x\in \R^n: \|x\|_{\dn}^{-\mu}+\mu\rho \tau\geq 0$ and $\forall \tau>0$, and $Q_{\tau}(\|x\|_{\dn})x=\zero$ otherwise.
Hence, we conclude 	\vspace{-2mm}
\[
\|Q_{\tau}(\|x\|_{\dn})x\|_{\dn}=(\|x\|_{\dn}^{-\mu}+\mu\rho \tau)^{-1/\mu}\leq \|x\|_{\dn}\leq \overline{\sigma}(\|x\|),	\vspace{-2mm}
\]
where $\overline \sigma\in \mathcal{K}_{\infty}$ is defined in Theorem \ref{thm:hom_norm}.
In this case,	there exists $\xi\in\mathcal{K}_{\infty}$ such that $|u(t_{k+j})|\leq \xi(\|x_{k+j}\|))$, $j=0,1,...,n-1$
and there exists $\sigma_{1}\in \mathcal{K}_{\infty}$ such that 
$
\|x_{k+j+1}\|_{\dn}\leq \sigma_{1}(\|x_{k+j}\|_{\dn}), \quad j=0,1,...,n-1.
$
Consequently, there exists $\sigma_{n}\in \mathcal{K}_{\infty}$ such that 
$
\|x_{k+n}\|_{\dn}\leq \sigma_{n}(\|x_{k}\|_{\dn}), \forall k\in \N.
$
On the other hand, by construction, we have  
$x_{k+n}=Q_{nh}(\|x_{k}\|_{\dn})x_{k}$, so, taking into account $\|Q_{nh}(\|x_{k}\|_{\dn})x_{k}\|_{\dn}\leq \|x_{k}\|_{\dn},$ we derive
$
\|x_{k+n}\|_{\dn}\leq \min\left\{\|x_{k}\|_{\dn}, \sigma_n(\|x_{k}\|_{\dn})\right\}
$
and
$
\|x_{k}\|_{\dn}\leq \min\left\{\|x_0\|_{\dn}, \sigma_n(\|x_{0}\|_{\dn}\right\},\forall k\geq 0.
$
Using Theorem \ref{thm:hom_norm} we {substantiated} the global Lyapunov stability of the system \eqref{eq:dicrete_model} (as well as the system \eqref{eq:main_system}) with the sampled-time control \eqref{eq:hom_consist_full}. 	

\subsection{The proof of Proposition \ref{prop:approx}}
First of all, notice that under Assumption \ref{as:1} we have $K_0=\zero$ and $A_0=A$ (see Theorem \ref{thm:hom_control}). 
Let us denote $s_h=\ln (1+\mu \rho nh r^\mu)^{1/(\rho \mu)}$  with $r=\|x\|_{\dn}$ and 
show $\tilde K_h(r) \to r^{1+\mu}K\dn(-\ln r)$ as $h\to 0$. 

On the one hand, if $\dn_*(s)$ is defined by Lemma \ref{lem:main} then \vspace{-2mm}
\[
\begin{array}{c}
	\dn_*\!(\text{--}\ln h)e^{(A+BK+\rho G_{\dn})s_h}\!=\!\!\sum\limits_{i=0}^{\infty}\!\tfrac{s_h^i	\dn_*\!(\text{--}\ln h)(A+BK+\rho G_{\dn})^i\!}{i!}\!=\\
	\sum\limits_{i=0}^{\infty}\tfrac{s_h^i (A+BK\dn_*(\ln h)+\rho h G_{\dn})^i \dn_*(-\ln h)}{h^i i!}=\\
	e^{\frac{s_h(A+\rho h G_{\dn})}{h}}\dn_*(-\ln h)+\sum_{i=1}^{n-1} \frac{s_h^i}{h^i i!}A^{i-1}BK+O(h).
\end{array}\vspace{-2mm}
\]
Indeed, for $i=2$ we have 
$$
(A+BK\dn_*(\ln h)+\rho h G_{\dn})^2 \dn_*(-\ln h)=
(A+\rho h G_{\dn})^2\dn_*(-\ln h)+ABK+ O(h),
$$
for $i=3$ we derive 
\[
(A+BK\dn_*(\ln h)+\rho h G_{\dn})^3 \dn_*(-\ln h)=
\]
\[%\scriptstyle
(A+BK\dn_*(\ln h)+\rho h G_{\dn}) \left\{(A+\rho h G_{\dn})^2\dn_*(-\ln h)+ABK+ O(h)\right\}=
\]
\[
(A+\rho h G_{\dn})^3\dn_*(-\ln h)+A^2BK+O(h)
\]
and, by induction,	we conclude 
$
%\begin{array}{c}
(A+BK\dn_*(\ln h)+\rho h G_{\dn})^i\dn(-\ln h)=
(A+\rho h G_{\dn})^i\dn_*(-\ln h)+A^{i-1}BK+O(h).
%\end{array}
$
Since $A$ is nilpotent then  for $i\geq n+1$ we have $A^{i-1}=\zero$ and $(A+BK\dn_*(\ln h)+\rho h G_{\dn})^i\dn(-\ln h)=
(A+\rho h G_{\dn})^i\dn_*(-\ln h)+O(h)$.

On the other hand, since $\dn(-s) A\dn(s)=e^{\mu s} A$ then $\dn(-s) e^{\tau A}\dn(s)=e^{e^{s\mu} \tau A}$ for all $s,\tau \in \R$ and 
$
Q_{nh}(r)-A_h^n=\dn( \ln r) \hat Q(s_h)\dn(-\ln r)-e^{nhA}=
\dn( \ln r)  (\hat Q(s_h)-e^{nhr^{\mu}A})	\dn(-\ln r)
$
where   $	\hat Q(s_h)$ is given by \eqref{eq:hatQ}.
%	\[
%	\hat Q(s_h)=e^{-\rho s_h G_{\dn} } e^{(A+BK+\rho G_{\dn})s_h}.
%	\]
Since 	$\dn^*(s)$ commutes with $\dn(\tau)$, $\forall s,\tau\in \R$ then using  the identities \eqref{eq:d_A_B} and the estimate of  
$\dn_*\!(\text{--}\ln h)e^{(A+BK+\rho G_{\dn})s_h}$ we derive \\
$
\begin{array}{c}
	\dn_*(-\ln h) \left(e^{(A+BK+\rho G_{\dn})s_h}-e^{\rho s_h G_{\dn}} e^{nhr^{\mu} A}\right)=\\
	\dn_*(-\ln h)e^{(A+BK+\rho G_{\dn})s_h}-e^{\rho s_h G_{\dn}} e^{nr^{\mu} A} \dn_*(-\ln h)=\\
	\sum\limits_{i=0}^{n-1} \!\!\frac{s_h^i\!A^{i\text{--}1}\!BK}{h^i i!}\!\!+\!\!\left(\!e^{\frac{s_h\!(A+\rho h G_{\dn})}{h}}\!\!-\!e^{\rho s_h\!G_{\dn}} e^{nr^{\mu}\!A}\!\right)\!\dn_*\!(\text{--}\!\ln h\!)\!+\!O(h).\\
\end{array}
$

If $Z_1=\rho s_h G_{\dn}$, $Z_2=nr^{\mu} A$, $q=-\mu\rho s_h $ then  the condition $Z_1Z_2-Z_2Z_1=q Z_2$ of  Lemma \ref{lem:exp_XY} is fulfilled, so 	$e^{Z_1}e^{Z_2}=e^{Z_1+\frac{q}{1-e^{-q}}Z_2}$ or, equivalently,
$
e^{\rho s_h G_{\dn}} e^{nr^{\mu} A} = e^{\rho s_h G_{\dn}+\frac{s_h}{h}A}.
$
Therefore, since 
$
s_h=%\ln (1+\mu \rho nh r^\mu)^{1/(\rho \mu)}=
\tfrac{\ln (1+\mu \rho nh r^\mu)}{\rho \mu}=\tfrac{\mu \rho nh r^\mu-\frac{(\mu \rho 2h r^\mu)^2}{2}+O(h^3)}{\rho \mu}=
hnr^{\mu}+O(h^2)
$
then 	
$
%	\begin{array}{c}
\dn_*(-\ln h) \left(e^{(A+BK+\rho G_{\dn})s_h}-e^{\rho s_h G_{\dn}} e^{nhr^{\mu} A}\right)=%\\
\sum_{i=0}^{n-1} \!\tfrac{s_h^i\!A^{i-1}\!BK}{h^i i!}+O(h)=\sum_{i=0}^{n-1} \!\tfrac{(nr^{\mu})^i\!A^{i-1}\!BK}{i!}+O(h)=	%\\
\int^{nr^{\mu}}_0 e^{\tau A} d\tau BK +O(h).
%	\end{array}
$
Since $\dn(s)=e^{sG_{\dn}}$ then\\    
\centerline{
	$
	\begin{array}{c}
		\dn_*(-\ln h) (Q_{nh}(r)-A_h^n)=\\
		\dn(\ln r)e^{-\rho s_h G_{\dn}}\int^{nr^{\mu}}_0 e^{\tau A} d\tau 	BK\dn(-\ln r)+O(h)=\\
		\int^{nr^{\mu}}_0 \dn(\ln r -\rho  s_h)e^{\tau A} d\tau 	BK\dn(-\ln r)+O(h)=\\
		\int^{nr^{\mu}}_0\!\!e^{e^{-\mu (\ln r -\rho  s_h)} \tau A} d\tau 	 \dn(\ln r -\rho  s_h) BK\dn(-\ln r)+O(h)=\\
		\int^{nr^{\mu}}_0e^{\frac{1+\mu \rho  n h r^{\mu}}{r^{\mu}} \tau A} d\tau 	\tfrac{r}{(1+\mu \rho n h r^{\mu})^{1/\mu}}BK\dn(-\ln r)+O(h)=\\
		\int^{nr^{\mu}}_0e^{\frac{1}{r^{\mu}} \tau A} d\tau 	B rK\dn(-\ln r)+O(h)=\\
		\int^{n}_0e^{ \tau A} d\tau 	Br^{1+\mu}K\dn(-\ln r)+O(h).
	\end{array}$
}
Therefore, using Lemma \ref{lem:main} and Remark  \ref{rem:Gd_diag} we conclude 
\vspace{-2mm}
\[
\begin{array}{c}
	\tilde K_h(r)=e_{n}^{\top} W_n^{-1} (Q_{nh}(s_h)-A_h^n)=\\
	e_{n}^{\top} W_n^{-1} \dn_*(\ln h) \left(\int^{n}_0\!\!e^{ \tau A} d\tau 	Br^{1+\mu}K\dn(-\ln r)+O(h)\right)\!=\\
	r^{1+\mu}K\dn(-\ln r)+O(h^{2}),
\end{array}	\vspace{-2mm}
\]
and $\tilde K_h(r)\to r^{1+\mu}K\dn(-\ln r)$ as $h\to 0$ uniformly on $r$ from compacts belonging to $(0,+\infty)$.

\subsection{The proof of Lemma \ref{lem:hom_z}}

Let us show that \eqref{eq:hom_z} holds.	Indeed, on the one hand, since $\|\dn(s)x\|_{\dn}=e^{s}\|x\|_{\dn}$ then 
\[
\hat Q\left(\tfrac{\ln (1+\mu
	\rho nhe^{-\mu s} \|\dn(s)x\|_{\dn}^{\mu})}{\rho \mu} \right)\!=\!
%	\]
%	\[
%	e^{
%		\frac{G_{\dn} \ln{1+\mu
%				\rho n he^{-\mu s} \|\dn(s)x\|_{\dn}^{\mu}}}{-\mu}} e^{\frac{(A+BK+\rho G_{\dn})\ln 1+\mu\rho nhe^{-\mu s} \|\dn(s)x\|_{\dn}^{\mu}}{\rho\mu}}=
%	\]
%\[	
\hat Q\!\left(\tfrac{\ln (1+\mu
	\rho nh \|x\|_{\dn}^{\mu})}{\rho \mu} \right)
\]
and $\forall s\in \R, \forall x\in \R^n$ we have 
$$
Q_{nhe^{-\mu s}}(\|\dn(s)x\|_{\dn})=
%	$$
%	$$
%	\dn(\ln \|\dn(s)x\|_{\dn})	\hat Q_{nhe^{-\mu s}}(\|\dn(s)x\|_{\dn})\dn(-\ln \|\dn(s)x\|_{\dn})=
%	$$
%$$
\dn(s)  Q_{nh}(\|x\|_{\dn})\dn(-s).
$$
On the other hand, using  Lemma \ref{lem:main} we derive
\[
W_{e^{-\mu s}h}=\dn_*(-\mu s) W_h, \quad B_{e^{-\mu s}h}=e^{-\mu s-s} \dn(s)B_{h},
\]
$$L_{e^{-\mu s}h} \dn(s)=\dn\left( s\right)L_h, \quad \dn\left( s\right)F_h=F_{e^{-\mu s}h}\dn\left( s\right),$$
for all $s\in \R$ and for all $h>0$, where the identities $\dn_*(\tau)=e^{\tau I_n+\tau G_0}$ and $\dn(s)=e^{sI_n-s\mu G_0}$ are utilized for the analysis $L_h$ in order to conclude that  $e^{-s-\mu s} \dn_*(\mu s)\dn(s)=I_n$, where $G_0$ is defined in Theorem \ref{thm:hom_control}.  Hence,
we derive \eqref{eq:hom_z}. The identity \eqref{eq:hom_tilde_u} can be obtained in the same way.

\subsection{The proof of Lemma  \ref{lem:hom_con_discr}}
\textbf{1) Local Finite-Time Stability for $\mu<0$.}
Let us  show that the matrix $F_h$ is nilpotent.
Notice that  $F_h$ can be rewritten as follows\vspace{-2mm}
\[
\begin{array}{c}
	F_h=\left(I_n-[\zero \;\; \zero \;\; ... \;\; \zero \;\; B_h]W_{h}^{-1}A_h^{n-1}\right)A_h=
	\\
	\left(A_h^{1-n}W_hW_{h}^{-1}A_h^{n-1}-[\zero \;\; \zero \;\; ... \;\; \zero \;\; B_h]W_{h}^{-1}A_h^{n-1}\right)A_h=\\
	A_h^{1-n}(W_h-[\zero \;\; \zero \;\; ... \;\; \zero \;\; A_h^{n-1}B_h])W_h^{-1}A_h^n=\\
	A_h^{1-n}[B_h \;\; A_hB_h \;\; ... \;\; A_h^{n-2}B_h \;\;\zero]W_h^{-1}A_h^n
\end{array}\vspace{-2mm}
\]
Since 
$
W_h^{-1}[A_hB_h \; \;... \;\; A_h^{n-1}B_h \;\;\zero]=[e_2 \;\; ... \;\; e_n \;\;\zero]
$
then \vspace{-2mm}
\[
\begin{array}{c}	F_h^2\!=\!
	A_h^{1\text{--}n}[B_h \; A_h\!B_h \; ... \; A_h^{n-2}\!B_h \;\!\zero][e_2 \; e_3 \; ... \; e_n \;\!\zero]W_h^{\text{--}1}\!A_h^n\!=
	\\
	A_h^{1\text{--}n}[A_h\!B_h \; ... \; A_h^{n-2}\!B_h \;\zero\; \zero ]W_h^{\text{--}1}\!A_h^n.
\end{array}\vspace{-2mm}
\]	
%	Using the above identities  we derive 
%	\[
%	F_h^3=A_h^{1\text{--}n}[A_h^2\!B_h \; ... \; A_h^{n-2}\!B_h \;\zero\; \zero ]W_h^{\text{--}1}\!A_h^n F_h=
%	\] 
%	\[
%	A_h^{1\text{--}n}[A_h^2\!B_h \; ... \; A_h^{n-2}\!B_h \;\zero\; \zero ][e_2 \; e_3 \; ... \; e_n \;\!\zero]W_h^{\text{--}1}\!A_h^n=
%	\]
%	\[
%	A_h^{1\text{--}n}[A_h^3\!B_h \; ... \; A_h^{n-2}\!B_h \;\zero\; \zero
%	\; \zero ]W_h^{\text{--}1}\!A_h^n.
%	\]
Continuing the same considerations  we derive 
$F_h^n=\zero$.
%\[
%F_h^n=A_h^{1\text{--}n}[\zero \; ...\; \zero] W_h^{\text{--}1}\!A_h^n=\zero.	
%\]
On the one hand, since for $\mu<0$ we have $Q(\|x\|_{\dn})=\zero $ if $\|x\|_{\dn}\leq (-\mu \rho n h)^{-1/\mu}$ then
the closed-loop system becomes linear $x_{k+1}=F_hx_k$ for $\|x_k\|_{\dn}\leq (-\mu \rho n h)^{-1/\mu}=\left (\hat h/ h\right)^{1/\mu}$.	
On the other hand,  the inequality $\|F_h^{i} x_0\|_{\dn}\leq \left (\hat h/h\right)^{1/\mu}$ is equivalent to
$
\left\|\dn\left( \tfrac{1}{\mu}\ln \tfrac{h}{\hat h}\right)F^{i}_h x_0\right\|=
\left\|F^{i}_{\hat h} \dn\left( \tfrac{1}{\mu}\ln \tfrac{h}{\hat h}\right) x_0\right\|\leq 1,
$
and the inequality $\|x_0\|_{\dn}\leq \underline{r}^-\left (\hat h/h\right)^{1/\mu} $ is equivalent to 
$\left\|\dn(-\ln \underline{r}) \dn\left( \tfrac{1}{\mu}\ln \tfrac{h}{\hat h}\right)x_0\right\|\leq 1.$	
Therefore, the inequality $\|F^{i}_{\hat h} \dn(\ln \underline r^{\text{--}} )\|\!\leq \!1$ yields
$\left\|F^{i}_{\hat h} \dn\left( \tfrac{1}{\mu}\ln \tfrac{h}{\hat h}\right) \!x_0\right\|\!\leq\! \left\|\dn(-\ln \underline{r}^-) \dn\left( \tfrac{1}{\mu}\ln \tfrac{h}{\hat h}\right)\!x_0\right\|$.  
Hence, we derive $\|x_i\|_{\dn}=\|F_h^{i} x_0\|_{\dn}\leq \left (\hat h/h\right)^{1/\mu}$ for $i=1,...,n-1$  provided that 
$\|x_0\|_{\dn}\leq \underline{r}^-\left (\hat h/h\right)^{1/\mu}$.			
Taking into account the nilpotence of $F_h$, the latter implies local Lyapunov stability of the closed-loop system and   the finite-time convergence of solutions to zero.

\textbf{2) Practical Finite-time Stability for $\mu<0$.}	
{	 The proof repeats the proof of Theorem \ref{thm:ISS}, the case 1) for  $q=\zero$ and 
	gives 	 $\tfrac{d\|x\|_{\dn}}{dt}\leq -0.5\rho \|x\|_{\dn}^{1+\mu}$ for all $x:  \|x\|_{\dn}^{-\mu}\geq \tilde r$. Using Lemma \ref{lem:hom_z} we derive  $\Omega^-$ for $h\neq \hat h$ with $\overline r^-\!=\!\tilde r^{-1/\mu}$, where $\tilde r$ is defined in the proof of Theorem \ref{thm:ISS}. }
%Without loss of generality, we select  $\tilde r>|\mu| \rho \hat h$.  

\textbf{3) Practical Fixed-time Stability for $\mu>0$}
Let us prove, now, the practical fixed-time stability. 
On the one hand, since, by Theorem \ref{thm:hom_control} the canonical homogeneous norm is a Lyapunov function of the system satisfying \vspace{-2mm}
\[
\frac{d}{dt}\|x(t)\|_{\dn}=-\rho \|x(t)\|_{\dn}^{1+\mu}, \vspace{-2mm}
\]
then 
$
\|x(t+nh)\|_{\dn}^{-\mu}=\|x(t)\|_{\dn}^{-\mu}+\mu \rho n h,
$
and for $\mu>0$ we have 
$
\|x(t+nh)\|_{\dn}<\left(\mu \rho n h\right)^{-1/\mu}=\left(\frac{\hat  h}{h}\right)^{1/\mu}
$
independently of $x(t)$.
On the other hand,  by Corollary \ref{cor:hom_solution}, we have  
$
x(t+nh)=Q_{nh}(\|x(t)\|_{\dn})x(t),
$
so $\|Q_{nh}(\|\tilde x\|_{\dn})\tilde x\|_{\dn}\leq \left(\frac{\hat  h}{h}\right)^{1/\mu}, \forall \tilde x\in \R^n.$
Since the right-hand side of the  system can be represented as follows \vspace{-2mm}
$$
z_h(x)=F_hx+L_h  Q_{nh}(\|x\|_{\dn}) x, \quad L_h:=B_he_n^{\top}W_h^{-1},\vspace{-2mm}
$$
then, for any $x_0\in \R^n$ the solution $x_k, k=0,1,2,...$ of the discrete-time system  \eqref{eq:dicrete_model}, \eqref{eq:hom_consist_reduced} 
satisfies \vspace{-2mm}
\[
\begin{array}{l}
	x_1=F_hx_0+L_hy_1, \\
	x_2=F_h^2x_0+F_hL_hy_1+L_hy_2, \\
	...\\
	x_n=F_h^nx_0+F_h^{n-1}L_hy_1+F^{n-2}_hL_hy_2+...+L_hy_n,
\end{array}\vspace{-2mm}
\]
where $y_i=Q_{nh}(\|x_{i-1}\|_{\dn})x_{i-1}$, $i=1,2,...$. Since the matrix $F_h$ is nilpotent then $F_h^n=\zero$ and \vspace{-2mm}
\[
x_k\!=\!F_h^{n-1}\!L_hy_{k-n+1}+F^{n-2}_h\!L_hy_{k-n+2}+...+L_hy_k, \forall k\!\geq\! n.\vspace{-2mm}
\]
Since {\scriptsize$\|y_i\|_{\dn}\!\leq\!  \left(\frac{\hat  h}{h}\right)^{\frac{1}{\mu}}\! \Leftrightarrow
	\left\|\dn\left(\!\tfrac{1}{\mu}\!\ln \tfrac{h}{\hat h}\right)\!y_i\right\|_{\dn}\!\leq\! 1  \Leftrightarrow  \left\|\dn\left( \!\tfrac{1}{\mu}\!\ln \tfrac{h}{\hat h}\right)\!y_i\right\|\!\leq\! 1 
	$}
then \vspace{-2mm}
\[\begin{smallmatrix}
	\left\|\dn\!\left(\!\tfrac{1}{\mu}\!\ln \tfrac{h}{\hat h}\!\right)\!x_k\right\|_{\dn}\!= \left\|\sum_{i=1}^{n} \!\dn\!\left(\! \tfrac{1}{\mu}\!\ln \tfrac{h}{\hat h}\!\right)\!F_h^{i-1}L_h\dn\!\left( \!\text{--}\tfrac{1}{\mu}\!\ln \tfrac{h}{\hat h}\!\right)\!v_i\right\|_{\dn}\!\!,
\end{smallmatrix}\vspace{-2mm}
\]
where $v_i=\dn\left( \tfrac{1}{\mu}\ln \tfrac{h}{\hat h}\right)y_{k+n-i}$. Taking into account 
$F_{\hat h}=\dn\left(\!\tfrac{1}{\mu}\ln \tfrac{h}{\hat h}\!\right)\!F_h\dn\left(\!-\tfrac{1}{\mu}\ln \tfrac{h}{\hat h}\!\right),$\;
$L_{\hat h}\!=\!\dn\left(\!\tfrac{1}{\mu}\ln \tfrac{h}{\hat h}\!\right)\!L_h\dn\left(\! -\tfrac{1}{\mu}\ln \tfrac{h}{\hat h}\!\right)\!$ and $\|v_i\|\!\leq\! 1$ we derive
$
\left\|\dn\left( \tfrac{1}{\mu}\ln \tfrac{h}{\hat h} \right)\!x_k\right\|_{\dn}\!< \bar r^+, \forall k\!\geq\! n.
$ 
%The latter inequality is equivalent to \eqref{eq:r_max+}.

\textbf{4) Local Asymptotic Stability for $\mu>0$.}
{Since 
	$\tfrac{d\|x\|_{\dn}}{dt}\leq -0.5\rho \|x\|_{\dn}^{1+\mu}$ for all $x:  \|x\|_{\dn}^{\mu}\leq \tilde r^{-1}$
	then for $\mu>0$ the closed-loop system is locally asymptotically stable. Using Lemma \ref{lem:hom_z} we derive  $\Omega^+$   with $\underline r^+=\tilde r^{-1/\mu}$.}

\subsection{The proof of Lemma \ref{lem:sol_discr}}
The symmetry proven by Lemma \ref{lem:hom_z} yields  $$M_{h}(\|x\|_{\dn})x=\dn(s)M_{e^{\mu s}h}(\|\dn(-s)x\|_{\dn})\dn(-s)x$$ for all $x\in \R^n$, for all $s\in \R$ and $\forall h> 0$.
Hence, for any $x_0\in \R^n\backslash\{\zero\}$  we have 
$
x_{1}\!=\!M_{h}(\|x_0\|_{\dn})x_0\!=\!\dn(\ln \|x_0\|_{\dn})\Theta_1(\|x_0\|^{\mu}_{\dn}\hat h,v_0)v_0.$ 
Since
$\|x_1\|_{\dn}=\|x_0\|_{\dn} \left\|\Theta_1(\|x_0\|^{\mu}_{\dn},v_0) v_0\right\|_{\dn}$
then \\
$
x_{2}=M_h(\|x_{1}\|_{\dn})x_1
\!=\!M_h(\|x_{1}\|_{\dn}) \dn(\ln \|x_0\|_{\dn})\Theta_1(\|x_1\|_{\dn}\hat h,v_0)v_0\\
=\dn(\ln \|x_0\|_{\dn}) M_{\|x_0\|_{\dn}^{\mu} \hat h}\left(\|x_1\|_{\dn}/\|x_0\|_{\dn}\right)\Theta_1\left(\|x_0\|_{\dn}^{\mu},v_0\right)v_0		\\
=\!\dn(\ln \|x_0\|_{\dn}) \Theta_2\left(\|x_0|_{\dn}^{\mu},v_0\right)v_0.$
Repeating the above considerations  we derive \eqref{eq:sol_hom_disc}.

\subsection{The proof of Theorem \ref{thm:consistency}}

The approximation property is proven by Proposition  \ref{prop:approx}. Let us prove the consistency of stability properties.  
If the discrete-time system \eqref{eq:disc_time_z} is globally uniformly 
finite-time stable for some $h>0$ then due to dilation symmetry (see Lemma \ref{lem:hom_z}) it is globally uniformly finite-time stable for any $h>0$, in particular, for $h=\hat h$.

\textit{Necessity.} Let us consider \textbf{the case $\mu<0$.}  
The uniformity of the finite-time stability  and Lemma \ref{lem:hom_con_discr} guarantee that there exists $k^*\geq 1$ such that for any $x_0: \|x_0\|_{\dn}\geq \underline{r}^-$  
we have  $\|x_{k^*}\|_{\dn}< \|x_0\|_{\dn}$. 
The latter means that 
$
\|x_0\|_{\dn} \|\Theta_{k^*}\left(\|x_0\|^{\mu},v_0\right)v_0\|_{\dn}<\|x_0\|_{\dn}
$
and \vspace{-2mm}
$$
\|\Theta_{k^*}\left(r^{\mu},v_0\right)v_0\|_{\dn}<1, \forall  r\geq \underline{r}^-, \forall v_0\in S.\vspace{-2mm}
$$	
Denoting $\delta=r^{\mu}$  for $\mu<0$ we derive  the inequality \eqref{eq:Theta<1}.  \textbf{The case $\mu>0$} can be treated similarly.

\textit{Sufficiency.}  	
Let us denote $r_*=(\overline r^-)^{\mu}$ for $\mu<0$ and 
$r_*=(\underline r^+)^{\mu}$ for $\mu>0$.
Let us consider the candidate Lyapunov function  $V:\R^n\mapsto[0,+\infty)$ defined as follows
\vspace{-2mm}
\begin{equation}\label{eq:V}
	V(x)=\left\{
	\begin{array}{lcc}
		\|\dn(-\mu^{-1}\ln r^*)x\|_{\dn}^{p} & \text{ if } & \|x\|^{\mu}_{\dn}\leq r^*, \\
		\|\dn(-\mu^{-1}\ln r^*)x\|  & \text{ if } & \|x\|^{\mu}_{\dn}\geq  r^*,
	\end{array}
	\right.		\vspace{-2mm}
\end{equation}
where $p=\beta$ if $\mu<0$ and $p=\alpha$ if $\mu>0$, where $\alpha, \beta$ are given by Theorem \ref{thm:hom_norm}. By construction, $V$ is positive definite, radially unbounded and globally Lipschitz continuous  with the Lipschtiz constant $L=\|\dn(-\mu^{-1}\ln r^*)\|$ and Lemma \ref{lem:V}. 

a) Let us show that $V$ is a Lyapunov function for $\|x_0\|_{\dn}^{\mu}\in[r_*,r^*]$. Since 	$(\delta,v)\mapsto \|\Theta_{k^*}(\delta,v) v\|_{\dn}$ is a continuous function on the compact $[r_*,r^*]\times S$ then using \eqref{eq:Theta<1}  we derive $
\gamma= \max_{\delta \in [r_*,r^*]}	\|\Theta_{k^*}(\delta,v) v\|_{\dn}<1
$	
and  for $\|x_0\|^{\mu} \in [r_*,r^*]$ we have $\|x_{k^*}\|_{\dn}\leq \gamma \|x_0\|_{\dn}$.
Since 
$
\|x_{k^*}\|_{\dn}\leq  \gamma \|x_0\|_{\dn} \Leftrightarrow 
\|\dn(s)x_{k^*}\|_{\dn}\leq \gamma \|\dn(s)x_0\|_{\dn}, \forall s\in \R
$
then $V(x_{k^*})\leq \gamma^{p} V(x_0)$ for $\|x_0\|_{\dn}^{\mu}\in[r_*,r^*]$.

b) Let us show that $V$ is a Lyapunov function for $\|x_0\|_{\dn}^{\mu}\leq r_*$. Since $\tfrac{d\|x\|_{\dn}}{dt}\leq -0.5\rho \|x\|_{\dn}^{1+\mu}$ for all $x:\|x\|_{\dn}^{\mu}\leq r_*$ (see the proof of Lemma \ref{lem:hom_con_discr}, case 2), then 
$
\|x_{1}\|_{\dn}\leq \gamma_* (\|x_0\|_{\dn}^{\mu})\|x_0\|_{\dn}, \forall x_0: \|x_0\|^{\mu}_{\dn}\leq r_*
$, or equivalently,
$
V(x_1)\!\leq\!\gamma_*^{p}(\|x_0\|_{\dn}^{\mu}) V(x_0),\forall x_0: \|x_0\|^{\mu}_{\dn}\leq r_*,
$
where $\gamma_{*}(s)=(1+0.5\rho\mu s)^{-1/\mu}$ with $s> 0$.
%	Since the sets $\Omega^-$ and $\Omega^+$ are invariant then 
%the inequality 	$\|x_0\|^{\mu}\leq r_*$ guarantees 
%	$
%	V(x_{k})\leq \left(V^{\frac{|\mu|}{p}}(x_0)-\tilde \gamma_* V^{\frac{\mu+|\mu|}{p}}(x_0)\right)^{\frac{p}{|\mu|}}, $ $\forall k\geq 1$. 	Moreover, there exists $W_*\in \mathcal{K}_{\infty}$ for $\mu>-\beta$ (resp. $W_*\in\mathcal{K}$ for $\mu=-\beta$) such that  
%	\[
%  \left(V^{\frac{|\mu|}{p}}(x_0)-\tilde \gamma_* V^{\frac{\mu+|\mu|}{p}}(x_0)\right)^{\frac{p}{|\mu|}}\!-V(x_0)\!\leq\! -W_*(V(x_0))
%	\]
%	for $\|x_0\|^{\mu}\in (0, r^*]$. Indeed,  $\mu>0$ this fact is obvious. For $\mu<0$ we have $p=\beta$.  Let us consider $-\beta<\mu<0$  and the function
% the function $g:[0,+\infty)\mapsto \R$ defined as follows
%	$
%	g(h)=\left(r^{\frac{|\mu|}{\beta}}-h\right)^{\frac{\beta}{|\mu|}}-r,
%	$
%	where $r=V(x_0)$ is a positive parameter. Notice that the inequality $\|x_0\|^{\mu}\leq r_*$  for the case $\mu<0$ 
%	 is equivalent to  $\|x_0\|_{\dn}\geq \overline{r}^-$, i.e., $V(\|x_0\|_{\dn})\geq \tilde r:=\left(\overline{r}^-(\underline{r}^-)^{\frac{1}{\mu}}\right)^{\beta}$. Moreover, the scalar $\tilde \gamma_*$ can be selected small enough to guarantee  $\tilde r>\tilde \gamma_*$. Using the mean value theorem we derive
%	$
%	g(h)\!=\!g(0)+\dot g(\theta)h\!=\!-|\mu|^{-1}\beta \left( r^{\frac{|\mu|}{\beta}}- \theta \right)^{\frac{\beta}{|\mu|}-1}\!\leq\! -W_*(r), \forall r\!\geq\! \tilde r
%	$
%	 for $\theta\!\in\![0,\tilde \gamma_*]$  and some $W_*\in\mathcal{K}_{\infty}$.	

c)	Let us show that $V$ is a Lyapunov function for $\|x_0\|_{\dn}^{\mu}\geq r^*$. If $\|x_0\|_{\dn}^{\mu}>r^*$ then for all $k\geq n$ we have 
$\|x_{k}\|=0$ if $\mu<0$ and $\|\dn(-\mu^{-1}\ln r^*)x_k\|<1$ if $\mu>0$ (see the proof of Lemma \ref{lem:hom_con_discr}). This means that there exist $\gamma^*\in (0,1)$ such that $
V(x_k)\leq \gamma^* V(x_0), \quad \|x_0\|^{\mu}>r^*.
$
Without loss of generality we may assume that $k^*\geq n$ (otherwise we just take $nk^*$ instead of $k^*$ in all above  considerations). 

d) Therefore, for any $\bar r_*\in (0,r_*)$ and for any finite $\bar r^*>r_*$ there exists $\bar \gamma\in (0,1)$ such that 		\vspace{-2mm}
$$
V(x_{k^*})\leq \bar \gamma V(x_0), \quad \forall x_0 : \bar r_*\leq \|x_0\|^{\mu}\leq  \bar r^*.		\vspace{-2mm}
$$  
%The latter means that $V$ is a global strict Lyapunov function of the discrete-time system, which guarantees its global uniform asymptotic stability. 
Taking into account local finite-time (resp., asymptotic) stability and practical finite-time  (resp., fixed-time)  stability proven by Lemma \ref{lem:hom_con_discr} 	 for $\mu<0$ (resp., $\mu>0$) 	we complete the proof.

\subsection{The proof of Theorem \ref{thm:ISS0}}	
In a discrete time, the system \eqref{eq:system_q0} can be rewritten as \vspace{-2mm}
$$x^q_{k+n}=Q_{nh}(\|x^q_k\|_{\dn})x^q_k+\sum_{i=0}^{n-1} A_h^{n-i}\tilde q_{k+i} \vspace{-2mm}$$
where   $\tilde q_k=\int^{h}_{0} e^{A(h-\tau)}q_p(t_k+\tau) d\tau$ is the sampled-time realization of the external perturbation, so $\{\tilde q_k\}\in\ell^{\infty}$ for any $h>0$.
Let $V$ be defined as in Lemma \ref{lem:V}.
Since 
$\|Q_{nh}(\|x\|_{\dn})x\|^{-\mu}_{\dn}=\|x\|_{\dn}^{-\mu}+\mu\rho nh$ for all $x\in \R^n: \|x\|_{\dn}^{-\mu}\!+\!\mu\rho nh\geq 0$ then 
%  $\|Q_{nh}(\|x^q_k+\hat q_k\|_{\dn})(x^q_k+\hat q_k)\|^{-\mu}_{\dn}=\|x^q_k+\hat q_k\|_{\dn}^{-\mu}+\mu\rho nh$
$
\left\|x^q_{k+n}\!-\!\bar q_k\right\|^{-\mu}_{\dn}\!=\!\|x^q_k\|_{\dn}^{-\mu}+\mu\rho nh,$ where $ 
\bar q_k=\sum_{i=0}^{n-1} A_h^{n-i}\tilde q_{k+i}.
$  Moreover, since $\bar q_k$ is uniformly bounded, then for $\mu>0$ it guarantees a practical fixed-time stability.
In this case, we derive\vspace{-2mm}
$$
\begin{smallmatrix}
	V(x_{k+n}^q)-V(x_{k}^q)=V(x_{k+n}^q)-V(x_{k+n}^q\!-\bar q_k)+V(x_{k+n}^q\!-\bar q_k)-V(x_{k}^q)\\
	\leq \|\bar q_k\|+V(x_{k+n}^q\!-\bar q_k)-V(x_k^q)=\|\bar q_k\|+\overline{\sigma}^{-1}(\|x_k^q-\bar q_k\|_{\dn})-V(x_k^q)\\
	= \|\bar q_k\|+W(V(x_k^q)),
\end{smallmatrix}\vspace{-2mm}$$
where $W(V)\!=\!\overline{\sigma}^{-1}(\overline\sigma(V)^{-\mu}\!+\!\mu\rho nh)^{-1/\mu}\!-\!V.$ For $\mu\!>\!-\beta$ we have $W\!\in\! \mathcal{K}_{\infty}$ and  $V$ is an ISS Lyapunov function\;\cite{JiangWang2001:Aut}.

\subsection{The proof of Theorem \ref{thm:ISS}}
In a discrete time, the system \eqref{eq:system_q} can be rewritten as
%	\begin{equation}\label{eq:discr_q}
%	x^q_{k+1}=M_h(\|x^q_k\|_{\dn})x_k^q+hq_k,
%	\end{equation}
\begin{equation}\label{eq:discr_q}
	x^q_{k+1}=A_hx^q_k+B_h\tilde K_h(\|x^q_k+\hat q_k\|)(x^q_k+\hat q_k)+\tilde q_k,
\end{equation}
where $\{\hat q_k\},\{\tilde q_k\}\in\ell^{\infty}$ for any $h>0$, $\hat q_k=q_m(t_k)$ is the sampled-time realization of the  measurement noise and  $\tilde q_k=\int^{h}_{0} e^{A(h-\tau)}q_p(t_k+\tau) d\tau$ is the sampled-time realization of the external perturbation. Denote $q_k=(\tilde q_k^{\top}, \hat q_k^{\top})$. 
Due to the dilation symmetry proven by Lemma \ref{lem:hom_z}  it is sufficient to analyze ISS of \eqref{eq:discr_q} for $h=\hat h$. 	 

\textbf{1)} Let us prove local ISS and practical ISS of \eqref{eq:system_q}.  If $x=e^{Ah_t}x^q_k+B_{h_t}\tilde u_{\hat h}(x^q_k+\hat q_k)+\tilde q_{h_t}$, $h_t=t-t_k$ and 
$\tilde q_{h_t}=\int^{h_t}_{0} e^{A(h_t-\tau)}q_p(t_k+\tau) d\tau$
then $x$ corresponds to a solution of the system \eqref{eq:system_q} for $t\in [t_{k},t_{k}+\hat h)$. 
Let us denote  $q_k^1=\dn(-\ln \|x^q_k\|_{\dn})\hat q_k$, $q^2_{k}\!=\!\dn(-\ln \|x^q_k\|_{\dn}) \tilde q_{h_t}$.\vspace{-2mm} 

a) Let us show that $\|x^q_k\|_{\dn}$ is close to $\|x\|_{\dn}$ for a sufficiently large $\|x^q_{k}\|^{-\mu}_{\dn}$ and sufficiently small $q_k^i$, $i=1,2$. 
Using dilation symmetry (see \eqref{eq:hom_tilde_u})  we derive \\
{\scriptsize$x\!=\!\dn(\ln \|x^q_k\|_{\dn})\!\left(\!e^{Ah_t\!\|x^q_k\|^{\mu}_{\dn}}v_k\!+\!B_{h_t\!\|x^q_k\|_{\dn}^{\mu}}\tilde u_{\hat h \|x^q_k\|^{\mu}}(\!v_k\!+\!q^1_k\!)\!+\!q^2_k\!\right)$\!,}
% 	$	\|x\|_{\dn}=\|x_i\|_{\dn}\left\|e^{Ah_t\|x_i\|^{\mu}_{\dn}}v_i+B_{h_t\|x_i\|_{\dn}^{\mu}}\tilde u_{\hat h}(v_i)\right\|_{\dn}$,
with $v_k=\dn(-\ln \|x^q_k\|_{\dn})x^q_k\in S$, 
Since $e^{Ah_ts}\to I_n $ as $s\to 0$ and $B_{h_t s}\to \zero$ as $s\to 0$ then 
for any $\varepsilon\in (0,1)$ there exist $r_{\varepsilon}>0$ and $\delta_{\varepsilon}>0$ such that \vspace{-2mm}
$$
\left|\left\|e^{Ah_t\|x^q_k\|^{\mu}_{\dn}}v_k\!+\!B_{h_t\|x^q_k\|_{\dn}^{\mu}}\tilde u_{\hat h}(v_k+q_k^1)+q_k^2\right\|_{\dn}\!\!-\!1\right|\!\leq\! \varepsilon\vspace{-2mm}
$$ 
for  $\|x^q_k\|^{-\mu}_{\dn}\!>\!r_{\varepsilon}$
and $\|q^i_{k}\|\leq \delta_{\varepsilon}, i=1,2$.
Hence, we have  $(1-\varepsilon)\|x^q_k\|_{\dn}\!\leq\! \|x\|_{\dn}\!\leq\! (1+\varepsilon)\|x^q_k\|_{\dn}$
and $\exists \tilde C_0,\tilde C_1\!>\!0: $
$\|\dn(\text{--}\ln \|x\|_{\dn})\hat q_k\|\!=\!\left\|\dn\left(\!\text{--}\ln \frac{\|x\|_{\dn}}{\|x^q_k\|_{\dn}}\!\right)q_k^1\right\|\!\leq\! \tilde C_{0} \|q^1_k\|$  	
and 
$\|\dn(-\ln \|x\|_{\dn})x^q_k\|<\tilde C_1$ for $\|x^q_k\|^{-\mu}_{\dn}\!>\!r_{\varepsilon}$ and  $\|q^i_{k}\|\leq \delta_{\varepsilon}$.  	 

b) Let us show that $u_{h}(x^q_k+\hat q_k)$ is close to $u(x)$  for a sufficiently large $\|x^q_k\|_{\dn}^{-\mu}$ and sufficiently small $\|q^i_{k}\|$. 
Using Lemma \ref{lem:main} and the identity $x^q_k=e^{-Ah_{t}}x-e^{-Ah_t}B_{h_t}\tilde u_{\hat h}(x^q_k+\hat q_k)-e^{-Ah_t}\tilde q_{h_t}$ we derive\vspace{-2mm}
$$
\begin{smallmatrix}
	\dn(-\ln \|x\|_{\dn})x-\dn(-\ln \|x\|_{\dn})(x_k^q+\hat q_k)=\\
	\dn(-\ln \|x\|_{\dn})(I_n-e^{-Ah_t})x+\dn(-\ln \|x\|_{\dn})e^{-Ah_t}B_{h_t}\tilde u_{\hat h}(x^q_k+\hat q_k)\\
	+\dn(-\ln \|x\|_{\dn})e^{-Ah_t}\tilde q_{h_t}+\dn(-\ln \|x\|_{\dn})\hat q_k=\\
	-\dn(-\ln \|x\|_{\dn}) \!\int^{h_t}_0\!\!Ae^{A\tau} d\tau	x+ 		
	e^{-Ah_t\|x\|_{\dn}^{\mu}}\dn(-\ln \|x\|_{\dn})B_{h_t}\tilde u_{\hat h}(x^q_k+\hat q^k)\\
	+	e^{-Ah_t\|x\|_{\dn}^{\mu}}\dn(-\ln \|x\|_{\dn})\tilde q_{h_t}+\dn(-\ln \|x\|_{\dn})\hat q_k=\\
	-\int^{h_t\|x\|^{\mu}_{\dn}}_0Ae^{A\tau} d\tau	\dn(-\ln \|x\|_{\dn})x\\
	+e^{-Ah_t\|x\|_{\dn}^{\mu}}\,\cdot\, B_{h_t\|x\|_{\dn}^{\mu}}\,\cdot\, \tilde u_{\|x\|^{\mu}_{\dn}\hat h}\left(\dn\left(-\ln (\|x\|_{\dn}/\|x^q_k\|_{\dn})\right)(v_k+q_k^1)\right) \\
	% 		 		 		 		 \end{smallmatrix}
	% 	$$
	% $$
	%\begin{smallmatrix}	
	+ 	e^{-Ah_t\|x\|_{\dn}^{\mu}}\dn\left(-\ln (\|x\|_{\dn}/\|x^q_k\|_{\dn})\right) q_k^2+\dn(-\ln \|x\|_{\dn})\hat q_k.\vspace{-2mm}
\end{smallmatrix}
$$ 
Since   
$\exists \tilde C_2>0: \left\|\int^{h_t\|x\|^{\mu}_{\dn}}_0Ae^{A\tau} d\tau \right\|\leq \tilde C_2\|x\|^{\mu}_{\dn}$ and 
$\exists \tilde C_3>0: \left\|B_{\tilde h \|x\|^{\mu}_{\dn}}\right\|=\left\|\int^{h_t\|x\|^{\mu}_{\dn}}_0e^{A\tau} d\tau B\right\|\leq \tilde C_3\|x\|^{\mu}_{\dn}$
then for any $\tilde C_4>0$ there exist $C_3>0$ such that 
$
\|\dn(-\ln \|x\|_{\dn})x-\dn(\text{--}\ln \|x\|_{\dn})(x_k^q+\hat q_k)\|\!\leq\! C_3\|x\|_{\dn}^{\mu}
$ for $\|x\|_{\dn}^{-\mu}\geq \tilde r_{\varepsilon}=(\frac{1+\varepsilon}{1-\varepsilon})^{|\mu|} r_{\varepsilon}$, $\|q_k^i\|\!\leq\! \delta_{\varepsilon}$  and $\|\dn(-\ln \|x\|_{\dn})\hat q_k\|\leq\tilde C_4 \|x\|_{\dn}^{\mu}$.
In this case, by Lemma \ref{lem:u_u_h} there exists $C>0$ and $r>0$
such that 
$\|\tilde u_{\hat h}(x_k^q+\hat q_k)-u(x)\|\leq C\|x\|^{1+2\mu}$.
for  $\|x\|_{\dn}^{-\mu}\geq \max\{\tilde r_{\varepsilon},r^{-1}\}$, all $t\in [t_i,t_{i+1})$ 	and $\|q_k^i\|\leq \delta_{\varepsilon}$ and $\|\dn(-\ln \|x\|_{\dn})\hat q_k\|\leq \tilde C_{4}\|x\|_{\dn}^{\mu}$.

c) %{\color{red}Let us prove a forward completeness for $\mu\in [-1,0)$.} 
Adding and subtracting $u(x)$ we derive \vspace{-2mm}
$$
\begin{smallmatrix}
	\tfrac{d\|x\|_{\dn}}{dt}= \tfrac{\|x\|_{\dn}x^{\!\top}\!\dn^{\!\top}\!(\text{--}\ln \!\|x\|_{\dn})P\dn(\text{--}\ln \|x\|_{\dn}) (Ax+B\tilde u_{\hat h}(x_k+\hat q_k)+q_p)}{x^{\top}\dn^{\top}(-\ln \!\|x\|_{\dn})PG_{\dn} \dn(-\ln \|x\|_{\dn})x}=\\
	\tfrac{x^{\!\!\top}\!\!\dn^{\!\top}\!\!(\text{--}\ln \!\|x\|_{\dn}\!)P(B(\tilde u_{\hat h}\!(x_k+\hat q_k)-u(x))+\|x\|_{\dn}\dn(\text{--}\ln\!\|x\|_{\dn}\!) q_p)}{x^{\top}\dn^{\top}(-\ln \|x\|_{\dn})PG_{\dn} \dn(-\ln \|x\|_{\dn})x}-\rho \|x\|_{\dn}^{1\!+\!\mu}\\
	\leq \beta  \|B(\tilde u_{\hat h}(x_k+\hat q_k)-u(x))\|+\beta\|x\|_{\dn}\|\dn(-\ln \|x\|_{\dn}) q_p\|-\rho \|x\|_{\dn}^{1+\mu}\\
	\leq \|x\|_{\dn}^{1+\mu}(\beta\|B\| C\|x\|^{\mu}_{\dn}-\rho)
	+\beta\|x\|_{\dn} \|\dn(-\ln \|x\|_{\dn}) q_p\| \vspace{-2mm}
\end{smallmatrix}
$$
that 
for $\|x\|_{\dn}^{-\mu}\geq r'=\max\{\tilde r_{\varepsilon},r^{-1}\}$,  $t\in [t_i,t_{i+1})$,
$\|q_k^i\|\leq \delta_{\varepsilon}$   and $\|\dn(-\ln \|x\|_{\dn})\hat q_k\|\leq \tilde C_4 \|x\|_{\dn}^{\mu}$.

d) Let us show that $\|\cdot \|_{\dn}$ is an ISS Lyapunov function (close to zero for $\mu>0$ and close to infinity for $-\beta<\mu<0$).
Let $\sigma(s)=s^{\mu}\overline{\sigma}^{-1}(s)$, where $\overline{\sigma}\in \mathcal{K}_{\infty}$ is given in Theorem \ref{thm:hom_norm} and $s\geq 0$. For $\|q_p\|\leq 0.25\beta^{-1} \rho  \sigma(\|x\|_{\dn})$ we derive 
$
\beta \|x\|_{\dn} \|\dn(-\ln \|x\|_{\dn})q_p\|\leq 	 0.25\rho \|x\|_{\dn}^{1+\mu}.
$ Notice that $\|q_m\|\leq \delta_{\varepsilon} \overline{\sigma}^{-1}(\|x\|_{\dn})$ implies 
$\|q_k^1\|\leq \delta_{\varepsilon}$, 
$\|q_m\|\leq  \delta_0 \sigma(\|x\|_{\dn})$ with $\delta_0=\tilde C_0^{-1}\left(\frac{1-\varepsilon}{1+\varepsilon}\right)^{|\mu|}$ implies  $C_0\| q^1_k\|\leq \left(\frac{1-\varepsilon}{1+\varepsilon}\right)^{|\mu|}\|x^q_k\|_{\dn}^{\mu}$
and  $\|\dn(-\ln \|x\|_{\dn})\hat q_k\|\leq \|x\|_{\dn}^{\mu}$ for $\|x\|^{-\mu}\geq  r'$. For a sufficiently small $\delta'>0$ the inequalities
$\|x^q_k\|_{\dn}^{-\mu}\geq r_{\varepsilon}$ (or, equivalently, $\|x\|_{\dn}^{-\mu}\geq \tilde r_{\varepsilon}$) and $\|q_p\|\leq \delta' \sigma(\|x\|_{\dn})$ imply
$\|q_k^2\|\leq \delta_{\varepsilon}$. 	

Selecting $\tilde r=\max\{r', 0.5\rho (\beta\|B\| C)^{-1}\}$   
we derive $	\tfrac{d\|x\|_{\dn}}{dt}\leq -0.25\rho \|x\|_{\dn}^{1+\mu}$ for  $\|x\|_{\dn}^{-\mu}\geq \tilde r$, $\|q_{p}\|\leq \tilde \delta \sigma(\|x\|_{\dn})$, $\|q_m\|\leq  \delta_0 \sigma(\|x\|_{\dn})$ and $\|q_{m}\|\leq \delta_{\varepsilon}\overline{\sigma}(\|x\|_{\dn})$, where $\tilde \delta=\min\{0.25\beta^{-1} \rho,\delta'\}$.  The function $\sigma$ belongs to the class $\mathcal{K}_{\infty}$ if $\mu>-\beta$.
Therefore, the system \eqref{eq:system_q} is practically ISS if $-\beta<\mu<0$ and locally ISS if $\mu>0$  even when the conditions of Theorem \ref{thm:consistency} do not hold.

e) Let us show that the system  \eqref{eq:system_q} is practically fixed-time stable and practically ISS for $\mu>0$. Since $F_{\hat h}^{n}=\zero$ (see the proof of Lemma \ref{lem:hom_con_discr})  then  \vspace{-2mm}
\[\scriptsize
\begin{array}{lll}
	x^q_1\!=\!F_{\hat h}x^q_0+L_{\hat h}y_1+\bar q_1,& &\bar q_1=\tilde q_0- L_{\hat h}A_{\hat h}^n \hat q_0,\\
	x^q_2\!=\!F_{\hat h}^2x^q_0\!+\!F_{\hat h}L_{\hat h}y_1\!+\!L_{\hat h}y_2\!+\!\bar q_1, &  &\bar q_2\!=\!F_{\hat h}\!\bar q_1+\!\tilde q_1\!-\! L_{\hat h}A_{\hat h}^n \hat q_1,\\
	...& &...\\
	x^q_n\!=\!F_{\hat h}^{n-1}\!L_{\hat h}y_1\!+\!...\!+\!L_{\hat h}y_n\!+\!\bar q_n, & & \bar q_n\!=\!F_{\hat h}\bar q_{n\text{--}1}\!\!+\!\tilde q_{n\text{--}1}\!\!-\! L_{\hat h}\!A_{\hat h}^n \hat q_{n\text{--}1},
\end{array}\vspace{-2mm}
\]
where $y_{i+1}\!=\!Q_{n\hat h}(\|x^q_{i}\!+\!\hat q_{i}\|_{\dn})(x^q_{i}\!+\!\hat q_{i})$, $i\!=\!0,..., n-1$.
Since for $\mu>0$ we have $\|y_{k+i}\|\leq 1$ (see the proof of Lemma \ref{lem:hom_con_discr}, case 3)  
and $\|\bar q_{i}\|\leq C\max\limits_{j=0,...,i-1}\|q_{j}\|$ for some $C>0$, then the system  \eqref{eq:system_q} with $\mu>0$ is practically fixed-time stable and practically ISS.

f) Let us show that the system \eqref{eq:system_q} is locally ISS for $\mu\geq -1$. For $\|x_k^q+\hat q_k\|_{\dn}\leq \underline{r}^-$ we have  $x^q_{k+1}=F_{\hat h}x^q_k+\tilde  q_k-L_{\hat h}A^n_{\hat h}\hat q_k$, where $F_h$ is a Schur stable nilpotent matrix. Since an asymptotically stable linear system is ISS with respect to additive perturbations then the system   \eqref{eq:system_q} is locally ISS.

\textbf{2)} Let us show that the system \eqref{eq:system_q}  is ISS for $\mu>-\beta$ provided that the unperturbed system is globally asymptotically stable. 	
Our goal is to show a discrete-time system: \vspace{-2mm}
\begin{equation}\label{eq:system_q_k*}
	x^q_{(p+1)k*}\!\!=\!\Xi (x^q_{pk^*}, q_{pk^*},...,q_{(p+1)k^*-1}), \;\; p\!=\!0,1,...\!\vspace{-2mm}
\end{equation}
which describes evolution of  \eqref{eq:discr_q} with the discrete step $k^*$, is ISS. The latter would imply  ISS of \eqref{eq:system_q}. Notice that the local and practical ISS of \eqref{eq:system_q} guarantees the local and practical ISS of the system \eqref{eq:system_q_k*}. Let $k^*\geq 1$, $V$, $\bar r_*$, $\bar r^*,\bar \gamma \in (0,1)$, $L=\|\dn(-\mu^{-1}\ln r^*)\|$ be defined as in the proof of Theorem \ref{thm:consistency}. 	Let $x_k\in \R^n$ denote a solution of the non-perturbed system with $x_0=x_0^q$. 	 %\vspace{-2mm} 		

a) 
Let us show there exists $\omega_{k^*}\in\mathcal{K}_{\infty}$: 	\vspace{-2mm}
$$\|x_{k^*}^q-x_{k^*}\|\leq \omega_{k^*}(\max\{\|q_{0}\|,...,\|q_{k^*-1}\|\})\vspace{-2mm}$$ 
for  $\bar r_*\leq \|x_0\|_{\dn}^{\mu}\leq \bar r^*$.
Since the system  \eqref{eq:system_q}   is practically ISS then for any $\tilde \sigma \in K_{\infty}$ there exists a compact  set $\tilde \Omega\subset \R^n$
such that $x^q_{i}\in \tilde \Omega$ for all $i\in 0,...,k^*$ provided that $\bar r_*\leq \|x_0\|_{\dn}^{\mu}\leq \bar r^*$
and $\max\limits_{j=0,...,k^*-1}\|q_{j}\|\leq \tilde \sigma(\|x_0\|_{\dn})$.  	 	 
Denoting $\bar q_{0}=\zero$ and 
$\bar q_k=F_h\bar q_{k-1}+\tilde q_k+L_h(Q_{n\hat h}(\|x_{k-1}\!+\!\bar q_{k-1}\!+\!\hat q_{k}\|_{\dn})(x_{k-1}\!+\!\bar q_{k-1}\!+\!\hat q_{k}) -Q_{n\hat h}(\|x_{k-1}\|_{\dn})x_{k-1})$ for $k\geq 1$ we derive
$x_k^q=x_k+\bar q_k$. Since the function $x\mapsto Q_{\tau}(\|x\|_{\dn})x$ is continuous on $\R^n$ then by
Heine-Cantor Theorem it is uniformly continuous on $\tilde \Omega$ and there exists $\omega_0\in \mathcal{K}_{\infty}$ such that $\|Q_{\tau}(\|z_1\|_{\dn})z_1-Q_{\tau}(\|z_2\|_{\dn})z_2\|\leq \omega_0(\|z_1-z_2\|)$ for all $z_1,z_2\in \tilde \Omega$ and \vspace{-2mm}
$$ \|\bar q_{1}\|\leq \|\tilde q_0\|+\|L_h\|\omega_0(\|\hat q_0\|).\vspace{-2mm}$$
Repeating the above consideration,  on $k^*$-th step, we derive that 
$\exists \omega_{k^*}\in\mathcal{K}_{\infty}$: 
$\|\bar q_{k^*}\|\leq \omega_{k^*}(\max\limits_{j=0,...,k^*-1}\|q_{j}\|)$ for  $\bar r_*\leq \|x_0\|_{\dn}^{\mu}\leq \bar r^*$
and $\max\limits_{j=0,...,k^*-1}\|q_{j}\|\leq \tilde \sigma(\|x_0\|_{\dn})$.\vspace{-2mm}

b) Since   $V(x^q_{k^*})=V(x^q_{k^*})-V(x_k)+V(x_k)\leq 
L\|\bar q_{k^*}\|+\bar \gamma V(x_0)\leq\tfrac{\tilde \gamma +1}{2}V(x_0^q)
$
for $\bar r_*\leq \|x_0\|_{\dn}^{\mu}\leq \bar r^*$ and $\|\bar q_{k^*}\|\leq \frac{1-\bar \gamma}{2L}V(x_0)$ then  
$V(x^q_{k^*})\leq\tfrac{\tilde \gamma +1}{2}V(x_0^q)$ for $\bar r_*\leq \|x_0\|_{\dn}^{\mu}\leq \bar r^*$ and  $\max\limits_{j=0,...,k^*-1}\|q_{j}\|\leq \min\left\{\tilde \sigma(\|x_0\|_{\dn}),\omega_{k^*}^{-1}\left(\frac{1-\bar \gamma}{2L}V(x_0)\right)\right\}$.
Taking into account the local and practical ISS proven above, the latter guarantees 
global ISS of \eqref{eq:system_q_k*} in the view of  \cite{JiangWang2001:Aut}. The proof is complete.

\subsection{The proof of Corollaries \ref{cor:con_full_MIMO} and \ref{cor:con_reduced_MIMO}}	
Denote $\delta_i(t)=\sum_{j=i+1}^{m} A_{ij}x_{i}(t)$,	where  $i=1,...,m$ is a number of subsystem in the system \eqref{eq:main_system}, \eqref{eq:sampled_con}, \eqref{eq:hom_consist_full_MIMO} and 	the matrices $A_{ij}$ are defined in the proof of Corollary \ref{cor:cascade}. By Theorem \ref{thm:con_full}, each subsystem  with $\delta_i=0$ is finite-time (for $\mu_i<0$) or nearly fixed-time (for $\mu_i>0$) stable. Moreover, it is forward complete\footnote{A system is forward complete if all its solutions are defined globally in the forward time.} if $\delta_{i}$ is uniformly bounded. 
\textbf{The case $\mu_i<0$}. Since the $m$-th subsystem is finite-time stable then $\exists T_m>0$ such that $\delta_{m-1}(t)=\zero$ for all $t\geq T_m$.  Considering subsequently the systems $m-1$, $m-2$,...,$1$ we conclude that the system \eqref{eq:dicrete_model}, \eqref{eq:hom_consist_full_MIMO} is finite-time stable.
\textbf{The case $\mu_i>0$}. Since the $m$-th subsystem is fixed-time stable then $d_{m-1}$ is uniformly bounded and $(m-1)$-th subsystem practically fixed-time stable (see Theorem \ref{thm:ISS0}), but the ISS property guarantees its global uniform asymptotic stability \cite{SontagTeel1995:TAC}.  Using  the cascade structure of the system  we complete the proof of Corollary \ref{cor:con_full_MIMO}. The proof of Corollary  \ref{cor:con_reduced_MIMO} is literally the same but it uses Theorems \ref{thm:consistency} and \ref{thm:ISS} instead of Theorem 4 and 6, respectively.

\bibliographystyle{plain}

\end{document}